\documentclass[useAMS,usenatbib]{mnras}
\usepackage{epsf}
\usepackage{amssymb}
\usepackage{graphicx}
\usepackage{eqnarray,amsmath}
\usepackage{bigfoot}
\usepackage{verbatim}
\usepackage[usenames]{color}

\newcommand {\bc}{\begin {center}}
\newcommand {\ec}{\end {center}}
\newcommand {\be}{\begin {equation}}
\newcommand {\ee}{\end {equation}}

\newcommand {\Chandra}{{\it Chandra }}
\newcommand {\XMM}{{\it XMM-Newton }}
\newcommand {\XIPE}{{\it XIPE }}
\newcommand {\IXPE}{{\it IXPE }}
\newcommand {\ATHENA}{{\it ATHENA }}
\newcommand {\LYNX}{{\it LYNX }}

\def\deg{$^{\circ}$}

\title[Sgr~A* and molecular clouds
]{Can Sgr A* flares reveal the molecular gas density PDF?}
\author[Churazov et al.]{E.~Churazov$^{1,2}$,  I.Khabibullin$^{1,2}$, R.~Sunyaev$^{1,2}$, G.~Ponti$^{3}$\\
$^{1}$ MPI f\"ur Astrophysik, Karl-Schwarzschild str. 1, Garching D-85741, Germany\\
$^{2}$ Space Research Institute, Profsoyuznaya str. 84/32, Moscow
  117997, Russia\\
  $^3$ MPI f\"ur extraterrestrische Physik, Giessenbachstrasse 1, Garching D-85748, Germany \\
}

\begin{document}

\date{Accepted .... Received ...}

\pagerange{\pageref{firstpage}--\pageref{lastpage}} \pubyear{2012}

\maketitle

\label{firstpage}

\begin{abstract}
Illumination of dense gas in the Central Molecular Zone (CMZ) by
powerful X-ray flares from Sgr~A* leads to prominent structures in the
reflected emission that can be observed long after the end of the
flare.  By studying this emission we learn about past activity of the supermassive black hole in our
Galactic Center and, at the same time, we obtain unique information on
the structure of molecular clouds that is essentially impossible to
get by other means. Here we discuss how X-ray data can improve our knowledge of both sides of the
problem. Existing data already provide: i) an estimate of the flare
age, ii) a model-independent lower limit on the luminosity of Sgr~A*
during the flare and iii) an estimate of the total emitted
energy during Sgr~A* flare. On the molecular clouds side, the data
clearly show a voids-and-walls structure of the clouds and can
provide an almost unbiased probe of the mass/density distribution of
the molecular gas with the hydrogen column densities lower than few $10^{23}\;{\rm cm^{-2}}$. For instance, the probability distribution function of the gas density $PDF(\rho)$ can be measured this way. Future high energy resolution X-ray missions will provide the information on the gas velocities, allowing, for example a reconstruction of the velocity field structure functions and cross-matching the X-ray and molecular data based on positions and velocities. 
\end{abstract}

\section{Introduction}
\label{sec:road}
While the supermassive black hole Sgr A* at the center of the Milky
Way is currently very dim \citep[e.g.,][]{2003ApJ...591..891B}, it has experienced powerful
flares of X-ray radiation in the recent past. The historical
records of these flares are revealed by reflected/reprocessed
radiation coming from dense molecular clouds \citep{1993ApJ...407..606S,1996PASJ...48..249K}. The imprints left by the
flares in spatial and time variations of the reflected emission
suggest that a powerful flare happened some hundred years ago \citep[see, e.g.,][for review]{2013ASSP...34..331P}. It lasted
less than several years \citep[e.g.,][Churazov et al., 2017a]{2013A&A...558A..32C} and Sgr A* was more than million times
brighter than today. The energetics of the flare can be provided by a
relatively modest tidal disruption event \citep[e.g.][]{2012MNRAS.421.1315Z,2014ApJ...786L..12G}, although many issues remain unresolved.  

Clearly, molecular clouds
offer us a convenient tool to study Sgr A*’s past history. At the same
time, the flare itself serves as an extremely powerful probe of molecular
gas. We, in particular, argue that the X-ray illumination opens a new way of measuring the properties of the
gas density and velocity distributions in molecular clouds. The probability distribution function\footnote{Throughout the paper we use ``Probability Distribution Function'' in place of a more common ``Probability Density Function'' to avoid confusion with gas density.} of the gas density $PDF(\rho)$ inside a molecular cloud is shaped by a complex interplay between supersonic turbulence, self-gravity, magnetic fields and stellar feedback - the key agents in the current ISM paradigm \citep[see, e.g.,][for reviews]{2004ARA&A..42..211E,2004RvMP...76..125M,2007ARA&A..45..565M,2016SAAS...43...85K}. 
As a result, it bears invaluable information on how this interplay actually proceeds and in which way it determines the dynamical state of the cloud \citep[e.g.,][]{2010A&A...512A..81F}. The $PDF(\rho)$ is believed to be intimately connected to the stellar initial mass function and the efficiency of star formation \citep[e.g.][]{2002ApJ...576..870P,2004RvMP...76..125M}. Molecular clouds in the CMZ are of particular interest in this context, since they evolve in a very specific environment of the Galactic Center, and also because their measured star formation efficiency appears to be an order of magnitude lower compared to the molecular clouds in the Galactic disk \citep{2013MNRAS.429..987L,2014MNRAS.440.3370K,2016A&A...586A..50G,2016arXiv161003499K, 2017arXiv170403572B}. Measuring the $PDF(\rho)$ over broad range of the scales is a key for clarifying the particular mechanism(s) responsible for the suppression of star formation, e.g. high level of solenoidally-forced turbulence \citep{2016ApJ...832..143F}, and also whether this is a generic property of such an environment or it differs from a cloud to cloud, e.g. as a result of the different orbital evolution histories \citep{2016arXiv161003502K}.
Similarly important is the information on the properties of the velocity field \citep[e.g.,][]{2002ApJ...569..841B,2007ApJ...665..416K}, which can be characterized, e.g., by its structure function. Unfortunately, direct measurement of the $PDF(\rho)$ or the velocity structure function based on molecular emission lines is hindered by the projection effects, although various (model-dependent)  de-projection techniques have been proposed and tested on the data \citep[e.g.][]{1998A&A...336..697S,2010MNRAS.405L..56B,2014Sci...344..183K}. Further complication comes from the fact that the commonly used gas tracers are sensitive to a certain range of densities determined by self-absorption and collisional de-excitation \citep{1991ASIC..342..155G}. Below we argue that this limitations for the density distribution can be lifted by using X-ray observations of clouds illuminated by the Sgr~A* flare. This can be done with the current generation of X-ray observatories.  Future X-ray observatories, including cryogenic bolometers and polarimeters,
will further boost our ability to conduct in-depth studies of
molecular gas, providing, in particular, a possibility to measure the structure  function of the velocity field (almost) free from the projection effects.

In the previous two papers Churazov et al., 2017a and 2017b, hereafter \citeauthor{C17a} and \citeauthor{C17b}, we have considered constraints on the age of the flare, reconstruction of the 3D distribution of molecular gas and the expected long term evolution of the reflected emission. Here we focus more on the internal structure of molecular clouds and also outline a roadmap for detail diagnostics of the molecular gas using current and future observatories. In lieu of the description of the paper structure we refer to Table~\ref{tab:sum}, which provides a summary of possible measurements and links to the relevant sections. In this paper we consider only the molecular clouds in the so-called ``Bridge'' and G0.11-0.11, as these clouds are currently the brightest in the reflected emission.

\begin{table*}
\begin{center}
  \caption{Road-map for probing Sgr~A* flare and molecular clouds with X-ray data.
    \label{tab:sum}
  }
  \begin{tabular}{l | l | l}
    \hline
    Parameter     & Methods \& [Instruments] & Section \\
    \hline
    \hline
    $t_{age}$     & Structure functions in time and space [\Chandra, \XMM (multiple observations)] &\\   
    & Equivalent width of the 6.4 keV line [\Chandra, \XMM]  & \S\ref{sec:age}\\
    & Polarization [future X-ray polarimetric missions, e.g., \XIPE or \IXPE] &\\
        \hline
        $\Delta t$     &  Structure functions in time and space [\Chandra, \XMM (multiple observations)]&\\   
                  & Smallest clouds [\Chandra (deep + multiple observations)]& \S\ref{sec:dt}\\   
          \hline
        $L_{X,min}$     & Brightest clouds [\Chandra (deep)]& \S\ref{sec:tot}\\   
          \hline
          $L_X(t)$     & Smallest clouds  [\Chandra (deep + multiple observations)]& \S\ref{sec:dt}\\             \hline
          $L_{X}\Delta t$   & X-ray images + molecular data [\Chandra, \XMM] & \S\ref{sec:tot}\\   
          \hline
 
          $dV/d\rho$     & Deep X-ray image [\Chandra (deep)]& \S\ref{sec:tomo}\\
          \hline
          $\rho(l,b,z)$   & X-ray images  [\Chandra, \XMM (multiple observations)]&\S\ref{sec:tomo}\\
          \hline
          $v_z(l,b,z)$     & High resolution X-ray spectroscopy  [Athena, Lynx]&\S\ref{sec:future}\\
                    \hline

          $\vec{v}(l,b,z)$     & Proper motions from masers&\S\ref{sec:future}\\
    \hline
\end{tabular}
\end{center}
\end{table*}

\section{Parameters of the Sgr~A* flare}
\label{sec:flare}

\begin{figure*}
\begin{minipage}{0.99\textwidth}
\includegraphics[trim= 0mm 0cm 4.5cm 0cm, width=1\textwidth,clip=t,angle=0.,scale=0.92]{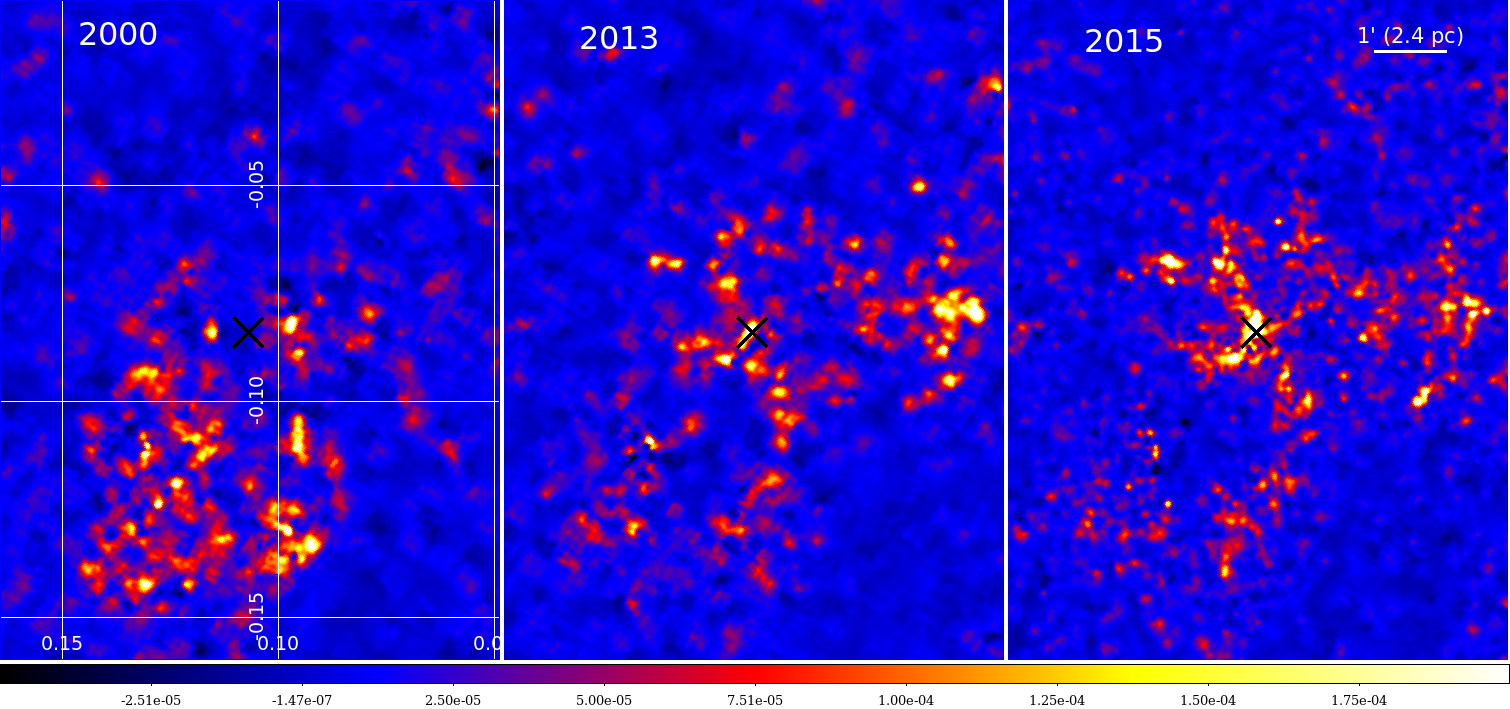}
\end{minipage}
\caption{\Chandra images of the reflected emission in a $7'\times9'$
  region centered at $(l,b)\sim(0.1,-0.08)$ in 2000, 2013 and
  2015. The images have been adaptively smoothed to get the same
  number of raw counts in the 4-8 keV band under a boxcar smoothing
  window. Images are shown in galactic coordinates. In a
  ``short-flare'' scenario, these images provide us a view of thin
  slices of gas ($\Delta z \sim 0.2 \Delta t_{yr}$~pc, where $\Delta t_{yr}$
  is the duration of the flare in years) through different layers
  of the gas. For observations separated by more than the duration of
  the flare, i.e. by several years, these slices do not overlap.  The
  2013 and 2015 images show a clear ``voids and walls'' structure of the
  molecular complex, resembling a typical outcome of numerical models \citep[see, e.g., Fig.\;9 in][]{2012MNRAS.424.2599C}. The earlier 2000 image gives an impression
  that the slice (in the bottom-left part of the image) is going
  through one of the ``walls'' of the molecular complex. Such slices are
  direct probes of the 3D density structure of this molecular complex. Black cross in all three images marks the brightest region in 2015 image.
\label{fig:tomo}
}
\end{figure*}
\begin{figure}
\includegraphics[trim= 6cm 4cm 5cm 2cm, width=1\textwidth,clip=t,angle=0.,scale=0.49]{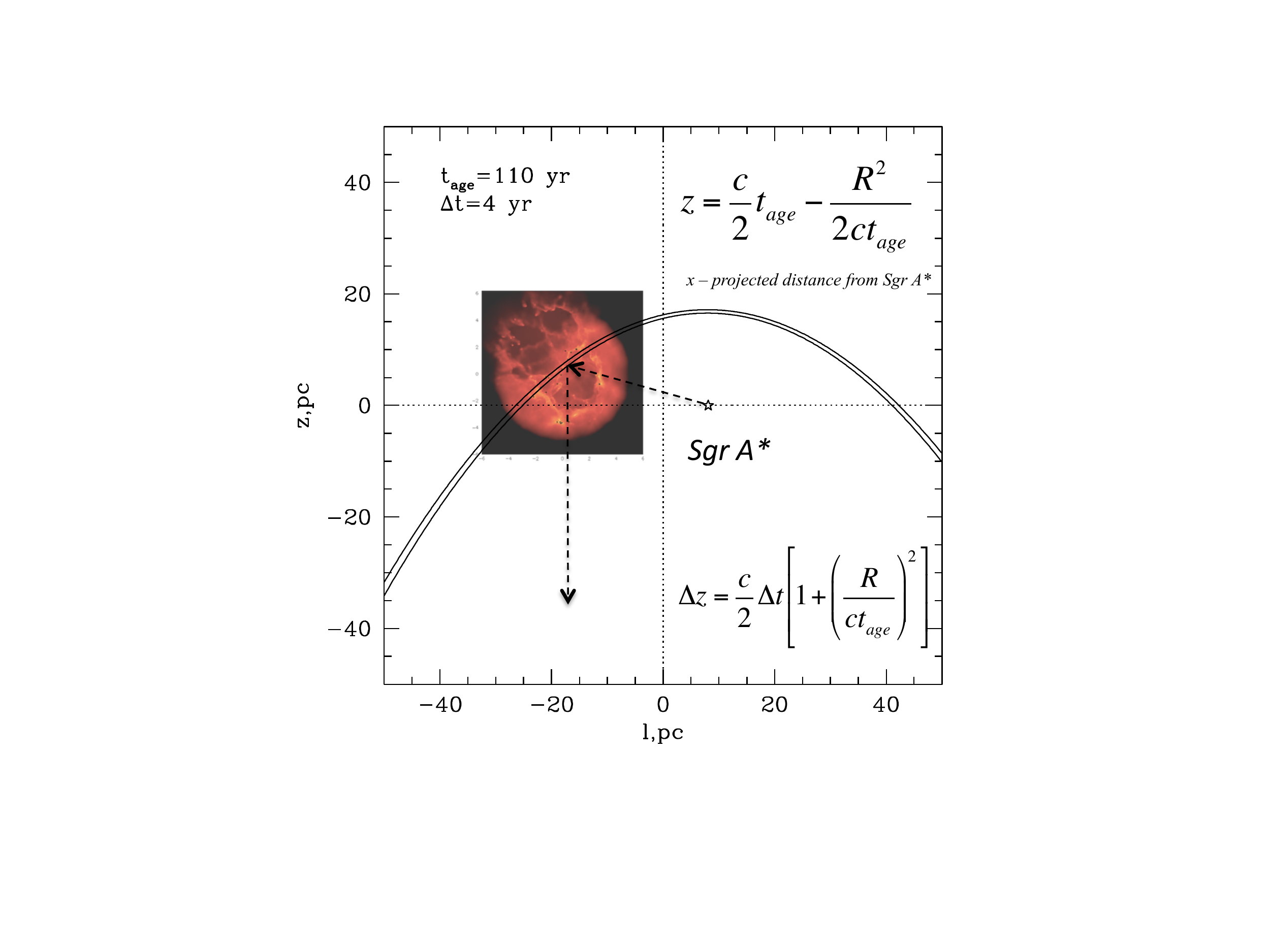}
\caption{Sketch of a molecular cloud exposed to a 4 years long flare
  that happened 110 years ago (a view on the Galactic Center region from above the Galactic Plane). An image of  a simulated cloud \citep[from][]{2012MNRAS.427..625W}  is used for illustration. Star marks the position of Sgr~A*. The light propagates from Sgr~A* to the cloud and then to observer.   The locus of illuminated gas is a space between two parabolas (see eq.\ref{eq:z}).  Only this gas should be visible in X-ray images. The thickness $\Delta z$ of the illuminated region along the line of sight is related to the duration of the flare $\Delta t$ and the projected distance $x$ from Sgr~A*.  The value $\Delta t=4$~yr is used only for illustration.
\label{fig:parabola}
}
\end{figure}

The reflected emission is formed as a result of interplay between three processes: Compton scattering, photoelectric absorption and fluorescence. For the energy band of interest here ($\sim$4-8~keV), there is no big differences between scattering by free electrons or electrons bound in hydrogen atoms or molecules unless the spectral resolution is high enough to resolve fine structure of the fluorescent lines caused by multiple scatterings \citep[][]{1996AstL...22..648S,1998AstL...24..271V,1999AstL...25..199S}. 
\Chandra images of the reflected emission from $7'\times9'$ region
centered at $(l,b)\sim(0.1,-0.08)$ taken in 2000, 2013 and 2015 are
shown in Fig.~\ref{fig:tomo} (see \citeauthor{C17a} for the details on the generation of
reflected emission maps). In short, the observed spectra in each pixel of an image  are decomposed into a linear combination of two components: one has a characteristic reflected spectrum with neutral iron 6.4\;keV line and another having  a spectrum of $\sim$6 keV optically thin plasma, to account for contribution of old stellar population. The normalizations of each component form two images.  From these images it is clear that the
reflected emission is highly structured and it varies strongly on
scales of 10 years or shorter. In addition, direct spectral
measurements of the variable emission show that its spectral shape is
consistent with the reflection scenario (see Fig.~4 in \citeauthor{C17b}).

The data on the reflected emission from molecular clouds provide us
several ways to probe the X-ray light curve $L_X(t)$ of Sgr~A* flare. As the first step one
can adopt a simplified model of the flare, by parameterizing it with
 three parameters - the age of the flare $t_{age}$, its duration
$\Delta t$ and luminosity $L_X$.  Of course, there are other
parameters affecting the problem, e.g., the shape of the flare
spectrum, abundance of heavy elements in the gas, etc., which are to be taken into account in the detailed quantitative model. In this section we first discuss the constraints that can be placed on   $t_{age}$, 
$\Delta t$, $L_X$ and assume that the flare emission is isotropic.

\subsection{Age of the flare}
\label{sec:age}
The age of the flare $t_{age}$ (the time elapsed since the onset
or end of the flare till the time of observation) is an important
parameter that is tightly linked to the position of the reflected
cloud with respect to the Sgr~A*. Indeed, the time delay arguments
\citep{1939AnAp....2..271C,1998MNRAS.297.1279S} imply that
\begin{eqnarray}
z=\frac{c}{2}t_{age}-\frac{R^2}{2\,c\,t_{age}},
\label{eq:z}
\end{eqnarray}
 where $z$ and $R$ are the line-of-sight and projected distances
of the cloud from Sgr~A*, respectively; $c$ is the speed of
light.  Since $R$ is directly measured, for a short flare there is a one-to-one correspondence
between $z$ and $t_{age}$. This is further illustrated in Fig.~\ref{fig:parabola} that schematically shows the locus of the illuminated gas for a short flare. 

There are several possibilities to measure $z$ or $t_{age}$, which we
discuss below.

\subsubsection{Comparison of time and space domains}
As discussed in \citeauthor{C17a}, the assumption that a small-scale substructure
of molecular clouds is isotropic in space allows one to determine the
age of the flare by comparing the structure functions of reflected
emission in time and space domains. If the flare is short, then even
from a single image, like one of those shown in Fig.~\ref{fig:tomo}, we have detailed
information on the space structure function of the illuminated gas (see \S\ref{sec:sf} below). By
observing the same region many times we get an information on the time
domain structure function. But for a short flare this time domain
structure function is essentially the space domain structure function,
subject to a simple transformation due to the velocity of the flare
propagation along the line of sight $v_z=\displaystyle \frac{\partial
  z}{\partial t}$. This value defines the shift $v_z\delta t$ of
the illuminated region along the line of sight between observation
separated by a given time $\delta t$. Comparing space and time domain
functions one can find $v_z$, which turns out to be $\sim 0.7\;c$
(\citeauthor{C17a}). The value of $v_z$ depends on the age of the flare
\begin{eqnarray}
v_z=\frac{c}{2}\left [1+ \left ( \frac{R}{c\;t_{age}}\right )^2\right].
\end{eqnarray}
Therefore, knowing $v_z$ one can determine the age of the
flare $t_{age}\sim 110\;{\rm yr}$ . While the accuracy of
this approach is limited by the amount of available data and rests on the assumption that on small scales the clouds are isotropic, the accuracy may improve in future, when more data become available. 

A more subtle diagnostics is possible from the elongation of illuminated clouds perpendicular to the direction towards Sgr~A* (see \citealt{1998MNRAS.297.1279S} and \citeauthor{C17a}), since the locus of illuminated region makes an angle to the line of sight, and this angle depends on the location of the cloud and   $t_{age}$. This should lead to the difference in structure functions calculated along radial (towards Sgr~A*) and tangential directions.

\subsubsection{Spectral analysis}

Another possibility to determine $z$ is to use the dependence of the
equivalent width of the iron line in the reflected emission on the
scattering angle \citep[][]{1998MNRAS.297.1279S}. Since the
fluorescent line is emitted isotropically, while the scattered
continuum intensity $\propto (1+\mu^2)$, the equivalent width varies
by a factor of two when the scattering angle changes from 0 to
90\deg. Therefore, fitting the spectra one can infer the scattering
angle. Such approach has been used in, e.g.,
\citet{2012A&A...545A..35C,2016MNRAS.463.2893W,2016arXiv161203320K}. The
accuracy of this approach is somewhat limited by the uncertainties in the heavy
elements abundance and the quality of modelling of other
background/foreground components. The latter limitation can be lifted by, e.g, considering the regions where reflected component dominates the spectrum, simultaneously fitting the spectra during and before/after the flare \citep{2013A&A...558A..32C}, or fitting only the variable part of the emission (\citeauthor{C17b}).

\subsubsection{Future polarization data}

A straightforward way of determining $z$ (and, therefore, $t_{age}$)
is possible with future X-ray polarization measurements
\citep[][C17b]{2002MNRAS.330..817C,2015A&A...576A..19M,2016A&A...589A..88M}. Effectively,
the degree of polarization $P$ is a proxy of the scattering angle
$\theta$.  More precisely, $\displaystyle P\approx
(1-\mu^2)/(1+\mu^2)$, where $\mu=\cos
\theta$. The uncertainty in the sign of $\theta$ can likely be
resolved based on the spectral data and the lightcurve. Once $\theta$
is known, the value of $z$ can immediately be derived from the
projected distance $R$ of the cloud from Sgr~A*. Knowing $R$ and $z$, the age of the
flare can be determined from eq.~(\ref{eq:z}). The sensitivity of
future X-ray polarimeters like \XIPE and \IXPE
\citep{2013ExA....36..523S,2013SPIE.8859E..08W} should be sufficient
to measure the degree of polarization and, therefore, $t_{age}$. More
importantly, a comparison of the polarization degree from different
clouds is a powerful test of the single-flare scenario that for a
given $t_{age}$ makes definite prediction of $P$ as a function of
projected distance from Sgr~A*. Furthermore, the direction of the polarization plane, which should be perpendicular to the direction to the primary source, is probably the most robust way to verify that for all clouds the source position is consistent with the location of Sgr~A*, in contrast with the assumption that individual clouds are illuminated by less luminous flares of stellar-mass X-ray objects local to these clouds \citep[see, e.g., a discussion of the reflected emission from the black hole binary 1E1740.7-2942 embedded in the cloud in][]{1991ApJ...383L..49S,1993A&AS...97..173C}.

\subsection{Duration of the flare}
\label{sec:dt}
Duration of the flare $\Delta t$ is one of the important parameters
that can be derived from observation. Below we briefly discuss two
ways of determining $\Delta t$. We assume below that there was only
one recent flare, unless explicitly stated otherwise.

\subsubsection{Light curves from individual clouds}
\label{sec:dt_d}
The simplest and the most direct way of constraining $\Delta t$ is by
studying the variability of a (small) molecular cloud in reflected
emission. Let us consider reflected emission coming from one
particular direction. Any given atom/electron is producing reflected
emission over time interval $\Delta t$ (if we neglect multiple
scatterings). A molecular cloud, occupying large region along the line
of sight will remain bright longer than $\Delta t$, because various
parts of the cloud are illuminated at different times. The time
scale of the illumination front propagation through the cloud is
$\Delta t_{c}\sim l_{z}/v_z$, where $l_{z}$ is the characteristic
size of the cloud along the line of sight\footnote{If we do not
  resolve the cloud, then one has to consider propagation time scales
  for all directions, which depend on the geometry of the cloud and its
  orientation with respect to the propagating front. The longest time scale
  will set the life time of the observed reflected emission caused by an
  infinitely short flare of the primary source.}  and $v_z$ depends on the position of the
cloud and the age of the flare $t_{age}$ (see \S\ref{sec:age}). For
example, in \citeauthor{C17a} we estimated $v_z\sim 0.7\,c$ for the Bridge
complex. From these arguments it is clear that (i) the shortest
observed flare places a conservative limit on $\Delta t$ and (ii)
systematic search of short flares for the smallest clouds is the best
strategy to provide the tightest constraints on the duration of the
flare. Moreover, if the cloud is small (i.e., $\Delta
t_{c}\ll\Delta t$) and relatively isolated, then the observed
light-curve from such cloud would be a direct probe of the Sgr~A*
light curve $L_X(t)$.

The variability of reflected emission has been modelled in many
studies
\citep[e.g.][C17b]{1998MNRAS.297.1279S,2014sf2a.conf...85C}. Observationally,
it was first reported in \cite{2007ApJ...656L..69M} and then found in
all regions bright in the reflected emission
\citep[e.g.][]{2010ApJ...714..732P,2010ApJ...719..143T,2012A&A...545A..35C,2013A&A...558A..32C,2013PASJ...65...33R,2015ApJ...814...94M,2015ApJ...815..132Z}.
 By now it is clear that Sgr~A* was variable on time scales
as short as several years (and, perhaps, on even shorter time
scales). More constraints will come soon from the multiple $\sim
100$~ks observations of the Bridge region with \Chandra (PI: Maica
Clavel). An ultimate set of constraints on the duration could come from
a set of very deep \Chandra observations that can probe the smallest
clouds (see \S\ref{sec:tomo}).

\subsubsection{Comparison of time and space domains}
\label{sec:dt_s}
Yet another possibility to determine $\Delta t$ was outlined in \citeauthor{C17a}.
Instead of studying individual clouds, one can compare statistical
properties of the reflected emission in space and time
domains. Basically, one calculates a correlation (or structure)
function for the reflected signal as a function of separation in space
and time and then compares these functions. As long as the
substructure in the gas density distribution is isotropic, one identify 
the effect of an extended flare (finite $\Delta t$) via its impact
onto the time domain correlation function. So far we did not see clear
signatures of finite duration (see \citeauthor{C17a}), but the data from ongoing
Chandra observations will likely have much better diagnostic power
than the existing data sets. 

The advantage of this approach is that all data are used to place
constraints on $\Delta t$ (rather than a light curve for a single
cloud). The approach, however, requires an assumption that there is
no difference in the statistics of small-scale structures along and
perpendicular to the line of sight. Also, to get unbiased results from the comparison of structure functions one has to take into account variations of the telescope angular resolution across the studied field, unless the images are heavily binned.

\subsection{Luminosity and total emitted energy}

\subsubsection{Lower limit on the luminosity of Sgr~A*}
\label{sec:min}

\begin{figure*}
\begin{minipage}{0.49\textwidth}
\includegraphics[trim= 0mm 4cm 0cm 2cm, width=1\textwidth,clip=t,angle=0.,scale=0.98]{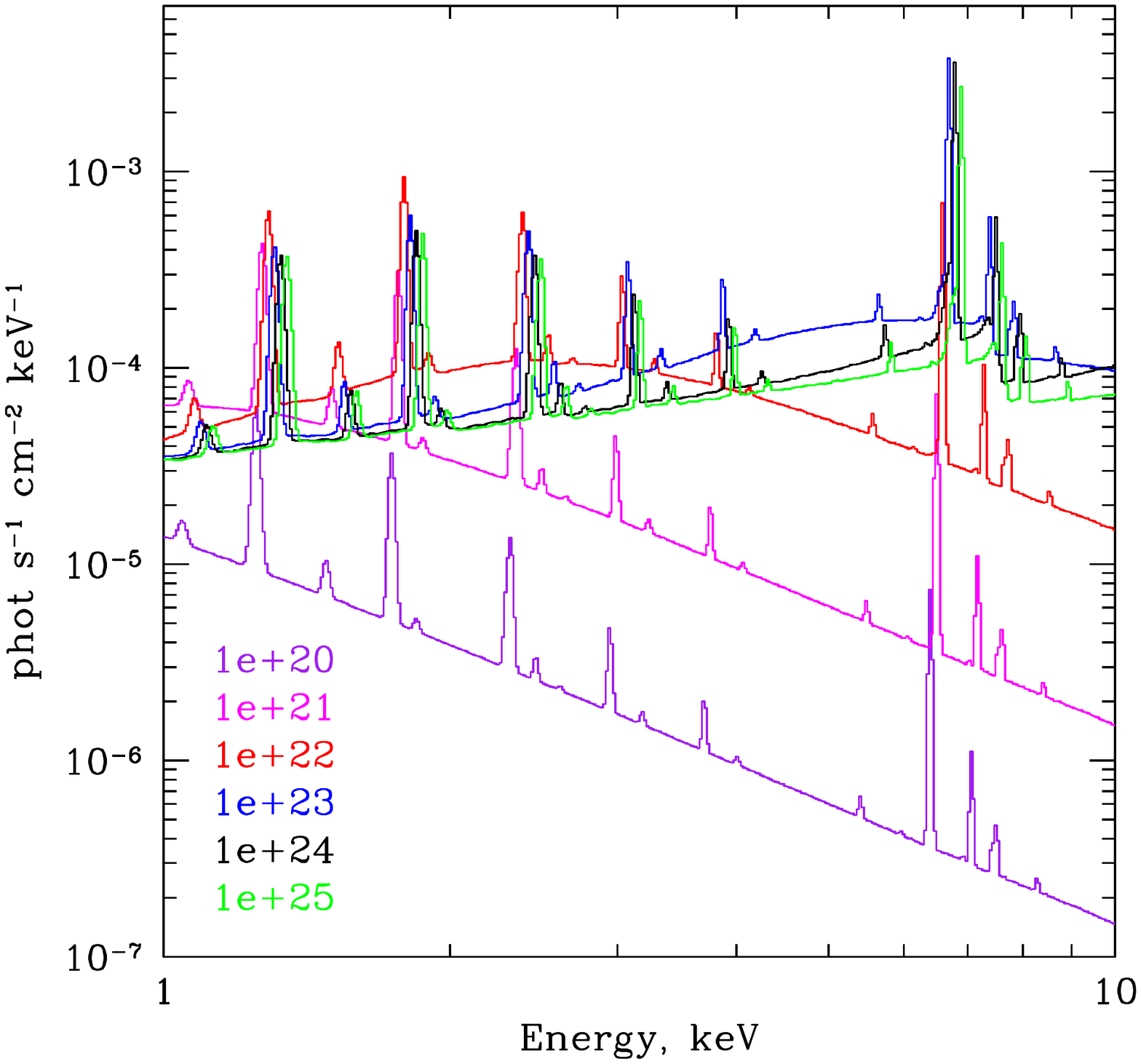}
\end{minipage}
\begin{minipage}{0.49\textwidth}
\includegraphics[trim= 0cm 4cm 0mm 2cm,width=1\textwidth,clip=t,angle=0.,scale=0.98]{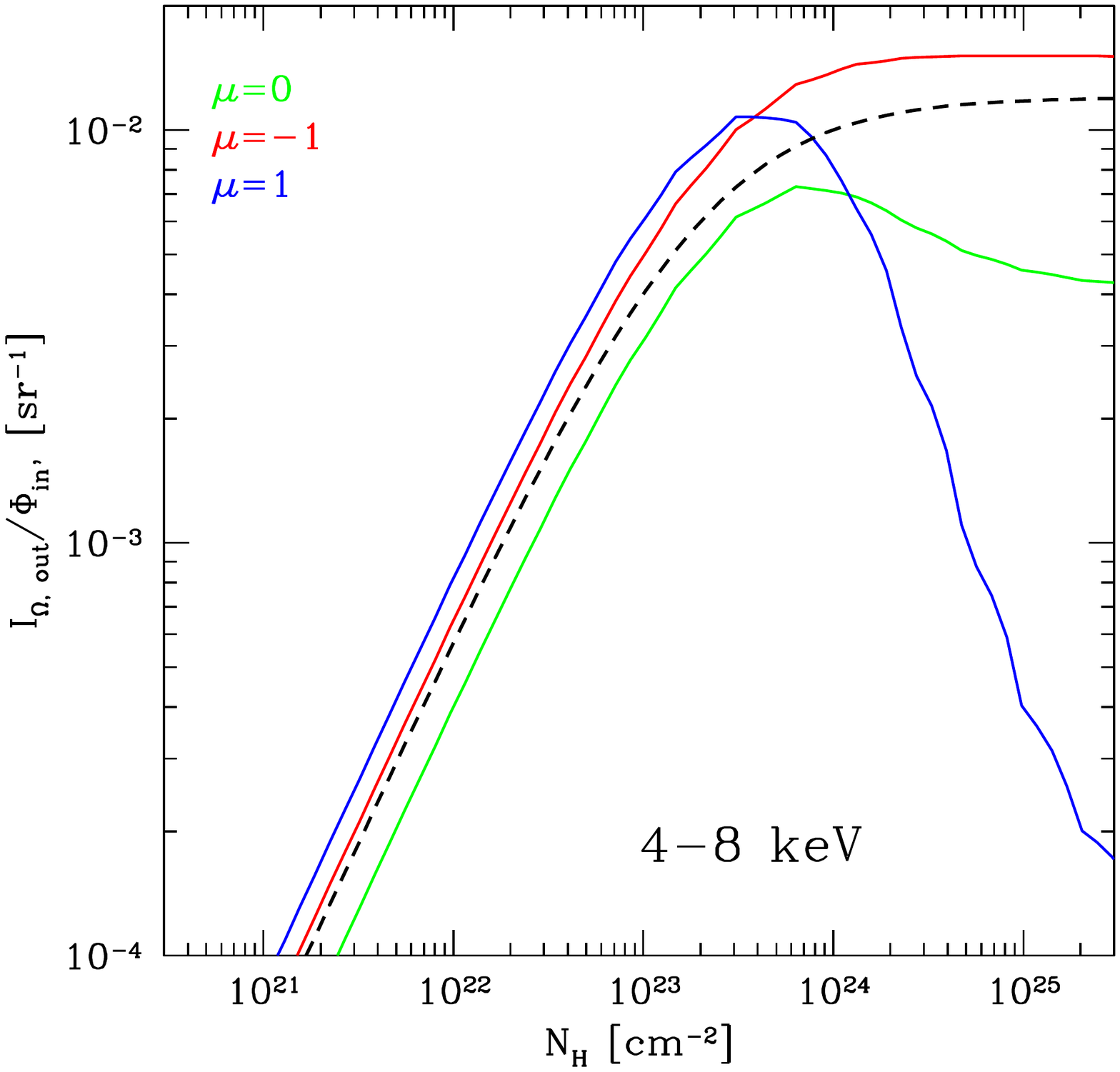}
\end{minipage}
\caption{Dependence of the reflected intensity on the column density
  and the scattering angle. {\bf Left:} Reflected spectra for a
  uniform spherical cloud of neutral gas illuminated by a parallel
  beam of X-rays with the power law spectrum with photon index
  $\Gamma=1.9$ for different values of the column densities $N_H$
  along the diameter of the cloud. The scattering angle is 90\deg. The
  {\small CREFL16} model (\citeauthor{C17b}) was used to generate the
  spectra. For small column densities the scattered continuum follows
  the shape of the incident spectrum and scales linearly with
  $N_H$. At large column densities the spectrum shape and
  normalization become independent on $N_H$. {\bf Right:} The ratio of
  radiant intensity to the incident flux in the 4-8 keV band as a
  function of $N_H$ for three scattering angles: 90\deg ($\mu=0$,
  green), 0\deg ($\mu=1$, blue) and 180\deg ($\mu=-1$, red). The
  dashed line is a crude approximation of this ratio (see
  eq.~(\ref{eq:eta}) in the text).
\label{fig:albedo}
}
\end{figure*}

Consider a uniform cloud of neutral gas, illuminated by a parallel
beam of X-ray emission with a power law spectrum with photon index $\Gamma=2$. Typical emergent
spectra are shown in Fig.~\ref{fig:albedo} (left panel) for the
scattering angle $\theta=$90\deg ($\mu=0$).  In all cases the
impinging energy flux $\Phi_{in}$ is the same, but the column density
of the cloud varies from $10^{20}$ to $10^{25}~{\rm cm^{-2}}$. From
Fig.~\ref{fig:albedo} it is clear that when the column density is
large, the efficiency of scattering becomes independent on
$N_{H}$. This happens at $N_{H}\sim 3\; 10^{23}~{\rm cm^{-2}}$, the
value that incidentally is not far from typical column densities for
molecular clouds in the CMZ \citep[e.g.][]{2017ApJ...835...76M}. For much larger column densities, the deepest
layers of the cloud are completely screened from the illuminating
radiation and do not contribute to the scattering signal.

We can characterize the efficiency of reflection  by
the quantity
\begin{eqnarray}
  \eta(N_H)=\frac{I_{\Omega,~out}}{\Phi_{in}}~{\rm sr^{-1}}
      \label{eq:eta}
\end{eqnarray}
where $I_{\Omega,~out}$ is the intensity of reflected emission in a
given direction (per unit solid angle), while $\Phi_{in}$ is the total
energy flux of X-ray radiation entering the cloud per unit time. The dependence of $\eta$ on $N_{H}$ and $\mu$ is shown in the right panel of Fig.~\ref{fig:albedo}. 
The dashed line in Fig.~\ref{fig:albedo} (left panel) shows the following crude approximation
\begin{eqnarray}
  \eta(N_H)\approx  1.2~10^{-2}\frac{N_{H,23}}{2+N_{H,23}} \lesssim 1.2~10^{-2},
      \label{eq:eta2}
\end{eqnarray}
where $N_{H,23}$ is the column density in units $10^{23}~{\rm
  cm^{-2}}$. Thus, for small column densities $\eta(N_H)\lesssim 6\;10^{-3}N_{H,23}$, while for very large column densities $\eta(N_H)\lesssim 1.2\;10^{-2}$.  
  The very fact that $\eta(N_H)$ does not exceed certain
value implies that one can use the maximal observed surface brightness
of the reflected emission to obtain a lower limit
on the luminosity of the illuminating source. To do this, we have
selected a small circular region with the radius of $2.8''$ centered
at the brightest spot in the 2015 images (see the black X in
Fig.~\ref{fig:tomo}). The spectrum extracted from this region has
strong fluorescent line at 6.4 keV and can be described as an almost
pure reflection component. Using the absorption-corrected 4-8 keV
surface brightness in the 4-8 keV band ($I_X\sim 6\,10^{-12}~{\rm
  erg\,s^{-1}\,cm^{-2}\,arcmin^{-2}}$) and the maximal value of
$\eta(N_H)\sim 10^{-2}$ we have estimated the incident flux
$\Phi_{in}$ in the same energy band. Finally, the luminosity of the
Sgr~A* was estimated by calculating the solid angle of the cloud as
seen from Sgr~A* position, $L_{X,\, 4-8\, keV}=\Phi_{in} 4 \pi
D^2_{sc}\sim 6\, 10^{38}\;{\rm erg\,s^{-1}}$, where $D_{sc}\approx
26~{\rm pc}$ is the assumed distance from the Sgr~A* to the
cloud (\citeauthor{C17a}). Recalculating 4-8 keV luminosity to other energy bands assuming
photon index of 2.0, we obtain the final result: the lower limit on
the luminosity $L_{X,min}$ of Sgr~A*, needed to generate observed
surface brightness is $\sim 10^{39}$ and $\sim 4\, 10^{39}~{\rm
  erg~s^{-1}}$ in the 2-10 and 1-100 keV bands, respectively. Such level of luminosity could in principle be provided by a stellar mass binary in the vicinity of the Galactic Center, rather than by Sgr~A* itself.  

Note that this is a conservative estimate, given that the conditions
required for maximal reflection efficiency, may not be satisfied in
the studied region. To reach the maximum efficiency the cloud has to be illuminated from the side facing the observer and the optical depth of the illuminated region along the line of sight has to be larger than $\sim3\; 10^{23}\;{\rm cm^{-2}}$.
 
Of course, other parameters, like the scattering angle, abundance of
heavy elements, etc., also affect the estimates of $L_X$. Variations
in these parameters lead to changes in the luminosity estimate by
factors of order 2, unless we consider extreme values of the
parameters.

\subsubsection{Total energy emitted by Sgr~A* during the flare}
\label{sec:tot}

Observations suggest that the reflected X-ray flux varies on scales as
small as few years \citep[e.g.][]{2013A&A...558A..32C}. In our
minimalistic model we assume that there was only one flare over the
last several hundred years. If so, the size of the illuminated slice along
the line of sight is $v_z\Delta t\sim 0.2\Delta t_{yr}$~pc, where $\Delta t_{yr}$ is
the duration of the flare in years (see fiducial parameters in
\S\ref{sec:age}). This value is much smaller than the characteristic size
of prominent molecular complexes, like, e.g., Bridge that has a
diameter of $\sim 10$~pc. Therefore, the flare illuminates only a thin
slice of the molecular complex. The mean density of this slice can be
estimated, assuming that it is the same as the mean density of the
entire complex.

To this end, we used the total H$_2$ mass of the GCM0.11~-0.08 cloud
$M\sim 1.3\,10^5~M_\odot$ from \citet{2017ApJ...835...76M} based on
the integrated line intensities in HCN and N2H+ (see their Figure
3). This mass corresponds to a rectangular region $5'\times 2'$ (or $12\times 5~{\rm pc}$). We
assume that the size of the molecular complex along the line of sight
is a geometric mean between the length and the width of the region,
i.e., $\sqrt{2\times 5}=3.2'$. The derived mean density (from the mass
and volume of the region) is $\bar{\rho}_H\sim 10^{4}~{\rm
  cm^{-3}}$. Thus, the mean column density of the illuminated layer $N_{H}\sim
\bar{\rho}_H v_z\Delta t\sim 7\,10^{21}~{\rm cm^{-2}}$. Therefore, the
Thompson depth of the layer is small and
the reflected flux scales linearly with the illuminating flux and the mass of the illuminated gas (see Fig.~\ref{fig:albedo}). Thus, the observed surface brightness of the reflected emission is
\begin{eqnarray}
I_X= \frac{L_X}{4\pi D_{sc}^2}\bar{\rho}_H v_z\Delta t \frac{\sigma_{4-8}}{4\pi},
\label{eq:ix_thin}
\end{eqnarray}
where $\sigma_{4-8}$ is the effective scattering cross section in the 4-8 keV band. For an optically thin medium, $\sigma_{4-8}$  can be evaluated from the value $\eta(N_H)$ by rewriting eq.~(\ref{eq:eta}) for the limit of $N_H\rightarrow 0$ as the ratio of the scattered flux to the illuminating flux entering a cloud with radius $r_c$ 
\begin{eqnarray}
\eta(N_H)=\frac{\frac{4}{3}\pi r_c^3 \bar{\rho}_H\frac{\sigma_{4-8} }{4\pi}\frac{L_X}{4\pi D_{sc}^2}}{\pi r_c^2 \frac{L_X}{4\pi D_{sc}^2}}=\frac{r_c \bar{\rho}_H}{3 \pi}\sigma_{4-8}=\frac{N_H}{6\pi}\sigma_{4-8},
\end{eqnarray}
where $N_H=2 r_c \bar{\rho}_H$ is the hydrogen column density of the cloud. The approximation given by eq.~(\ref{eq:eta2}) (see dashed line in Fig.~\ref{fig:albedo}) implies 
\begin{eqnarray}
\sigma_{4-8}\sim \frac{6\pi\eta(N_H)}{N_H}\sim 1.1\,10^{-24}~{\rm cm^{-2}}\approx 1.7 \sigma_T.
\label{eq:s48}
\end{eqnarray}
It is larger than the Thomson cross section per hydrogen atom due to (i) the presence of electrons in helium and other heavy elements, (ii) enhanced amplitude of the coherent scattering by multiple electrons in atoms heavier than hydrogen, and (iii) the contribution of the iron 6.4 keV line to the flux in the 4-8 keV band. For the molecular complex studied here, the estimate of the light front propagation velocity $v_z\sim 0.7 c$ (\citeauthor{C17a}) implies the cosine of the scattering angle $\mu\sim-0.4$. For such scattering angle the cross section is slightly lower due to the angular dependence of Compton scattering $\propto (1+\mu^2)$ and the values of $\eta(N_H)$ are close to the case of $\mu=0$ (see Fig.~\ref{fig:albedo}). Below we assume $\sigma_{4-8}\approx 1.4 \sigma_T$ to relate the X-ray surface brightness, the gas density and the luminosity of the primary source. 
The surface brightness of the same region measured by \Chandra during 2015 observations is $I_X\sim 4.3\,10^{-13}~{\rm erg\,s^{-1}\,cm^{-2}\,arcmin^{-2}}$ in the 4-8 keV band.
Assuming $v_z=0.7\,c$ the total energy emitted by Sgr A* during the flare is:
\begin{eqnarray}
  L_X\Delta t\sim \left ( \frac{\bar{\rho}_H}{10^4~{\rm cm^{-3}}} \right )^{-1}  \left\{\begin{array}{ll}
  3\,10^{46}~{\rm erg;} & 4-8~{\rm keV} \\ 
  6\,10^{46}~{\rm erg;} & 2-10~{\rm keV} \\ 
  2\,10^{47}~{\rm erg;} & 1-100~{\rm keV}, \\
\end{array}
\right. 
\label{eq:fluence}
\end{eqnarray}
where conversion from 4-8~keV to other bands is done assuming photon
index $\Gamma=2$. This is consistent with our previous estimates (\citeauthor{C17a}), but
now better justified value of mean density $\bar{\rho}_H\sim 10^4~{\rm cm^{-3}}$ is
used instead of $\bar{\rho}_H\sim 10^3~{\rm cm^{-3}}$. Note, that above estimate ignores the absorption and
scattering due to molecular gas on the photon path from the primary
source to the illuminated layer and then to the observer. Typical
absorbing column density in the Galactic Center region is few
$10^{22}~{\rm cm^{-2}}$, which has small impact on the 4-8~keV
flux. However, the column density of the molecular cloud itself is
$\sim \bar{\rho}_H \times 10~{\rm pc}\sim 3\,10^{23}~{\rm cm^{-2}}$. Such
column density would reduce the observed 4-8 keV flux by a factor of
$\sim 2$, increasing the estimate of the emitted power in
eq.~(\ref{eq:fluence}) accordingly.

We can now combine eq.~(\ref{eq:fluence}) with the lower limit
$L_{X,min}$ on the Sgr~A* luminosity (see \S\ref{sec:min}) to obtain
an upper limit on the duration of the flare:
\begin{eqnarray}
  \Delta t_{max} \sim \frac{L_X\Delta t}{L_{X,min}}\sim
  1.6~{\rm yr}.
  \label{eq:dtmax}
\end{eqnarray}
This value is consistent with constraints obtained so far from the
light curves of individual clouds \citep[e.g.][]{2013A&A...558A..32C}.
Shorter flares can not be excluded, unless one increases the estimate
of $L_{X,min}$ (see \S\ref{sec:min}) or finds very compact cloud with
the light crossing time  much shorter than a
year (see \S\ref{sec:dt_d}). For instance, for the Eddigton-level luminosity of $\sim
10^{44}~{\rm erg\,s^{-1}}$, the duration of the flare can be as short
as one hour, setting aside a question of what mechanism is responsible for such short flare.

\section{Tomography of molecular clouds}
\label{sec:tomo}
As discussed above, the apparent light curve of reflected emission in any given direction is a convolution of the intrinsic Sgr~A* light curve with the gas density distribution along the line of sight. As a result, the  observed light curve may a) appear longer than the flare itself and b) may possess a complicated substructure even if the intrinsic flare lacks it. Moreover, in \S\ref{sec:tot} we argue that the flare can be shorter than 1.6~yr (see eq.~\ref{eq:dtmax}), although this conclusion depends on the assumed mean density of the gas. Throughout the rest of the paper we consider the case of a short flare.

\subsection{Probability Distribution Function of Gas Density}

If the estimate of the duration $\Delta t\sim 1.6\;{\rm yr}$ is approximately correct, than the physical size of the illuminated region along the line of sight is $\Delta z\sim 0.7 c \Delta t\sim 0.2\; {\rm pc} \times \frac{\Delta t}{1\;{\rm yr}}$. At the same time, the arcsecond angular resolution of \Chandra corresponds to scales $\Delta x\sim 0.04\;{\rm pc}$\footnote{Of course, for off-axis angles larger than few arcminutes, the angular resolution of  \Chandra degrades significantly.}. Thus, X-ray observations give us a view of a narrow (less than 0.2~pc) density slice of the molecular cloud. Such data  provides a direct probe of the density distribution that is not contaminated by projection effects. In particular, it should be possible to reconstruct the Probability Distribution Function of the gas density $PDF(\rho)$. Indeed, according to eq.~(\ref{eq:ix_thin}) the surface brightness is proportional to the gas density, with the coefficient of proportionality set by the luminosity of the illuminating source, duration of the flare and geometrical factors. We note here that apart from the possible impact of the finite opacity, the relation between the density and the surface brightness is not sensitive to the physical state of the gas, i.e., the total mass density can be measured.   

Numerical simulations and analytic arguments suggest that supersonic  turbulence in the isothermal gas results in a log-normal $PDF(\rho)$ \citep{1994ApJ...423..681V,1997MNRAS.288..145P,1998ApJ...508L..99S,1999ApJ...524..169M,2001ApJ...546..980O, 2007ApJ...665..416K} with the characteristic width being determined by the Mach number of turbulent motions \citep{1997MNRAS.288..145P,1998PhRvE..58.4501P,2007ApJ...658..423K,2008ApJ...688L..79F,2012MNRAS.423.2680M}. The corresponding density variance-Mach number scaling is in turn sensitive to whether the turbulence is mainly compressively (e.g. via accretion or shocks) or solenoidally (e.g. via externally induced shear) driven \citep{2008ApJ...688L..79F}, and also on whether the gas is isothermal or not \citep{2015MNRAS.451.1380N}. As the cloud evolves, the self-gravity leads to the formation of power-law tails in the high density part of the density distribution \citep[e.g.,][]{1999ApJ...513..259O,2000ApJ...535..869K,2008ApJ...688L..79F,2011MNRAS.410L...8C,
2011ApJ...727L..20K,2013ApJ...763...51F,2014ApJ...781...91G,2017ApJ...834L...1B}.  

This high density tail is a key ingredient for determining the star formation rate and the star formation efficiency 
\citep[e.g.][]{2005ApJ...630..250K,2008ApJ...672.1006E,
2011ApJ...743L..29H,2011ApJ...730...40P,2015ApJ...806L..36S}.
Molecular clouds in the CMZ are of particular interest in this context, since they evolve in a very specific environment of the Galactic Center, and also because their measured star formation efficiency appears to be an order of magnitude lower compared to the molecular clouds in the Galactic disk \citep{2013MNRAS.429..987L,2014MNRAS.440.3370K,2016A&A...586A..50G,2016arXiv161003499K, 2017arXiv170403572B}. Measuring the $PDF(\rho)$ over broad range of the scales is a key for clarifying the particular mechanism(s) responsible for the suppression, e.g. high level of solenoidally-forced turbulence \citep{2016ApJ...832..143F}, and also whether this is a generic property of such an environment or it differs from a cloud to cloud, e.g. as a result of the different orbital evolution histories \citep{2016arXiv161003502K}. 

Thus, by measuring  $PDF(\rho)$ one gets insights on the driving mechanisms and the efficiency of star formation in a cloud. While the direct measurements of the $PDF(\rho)$ from molecular emission lines is hindered by the projection effects, there is a hope that due to thinness of the illuminated slice X-ray data could be free from this limitation. Another potential advantage of X-rays is their high penetrating power and insensitivity of the reflection efficiency to the physical state of the gas, unlike commonly used gas tracers that are sensitive to a certain range of densities determined by self-absorption and collisional de-excitation \citep[e.g.][]{1991ASIC..342..155G}. 

\begin{figure*}
\begin{minipage}{0.49\textwidth}
\includegraphics[trim= 0mm 4cm 0cm 2cm, width=1\textwidth,clip=t,angle=0.,scale=0.98]{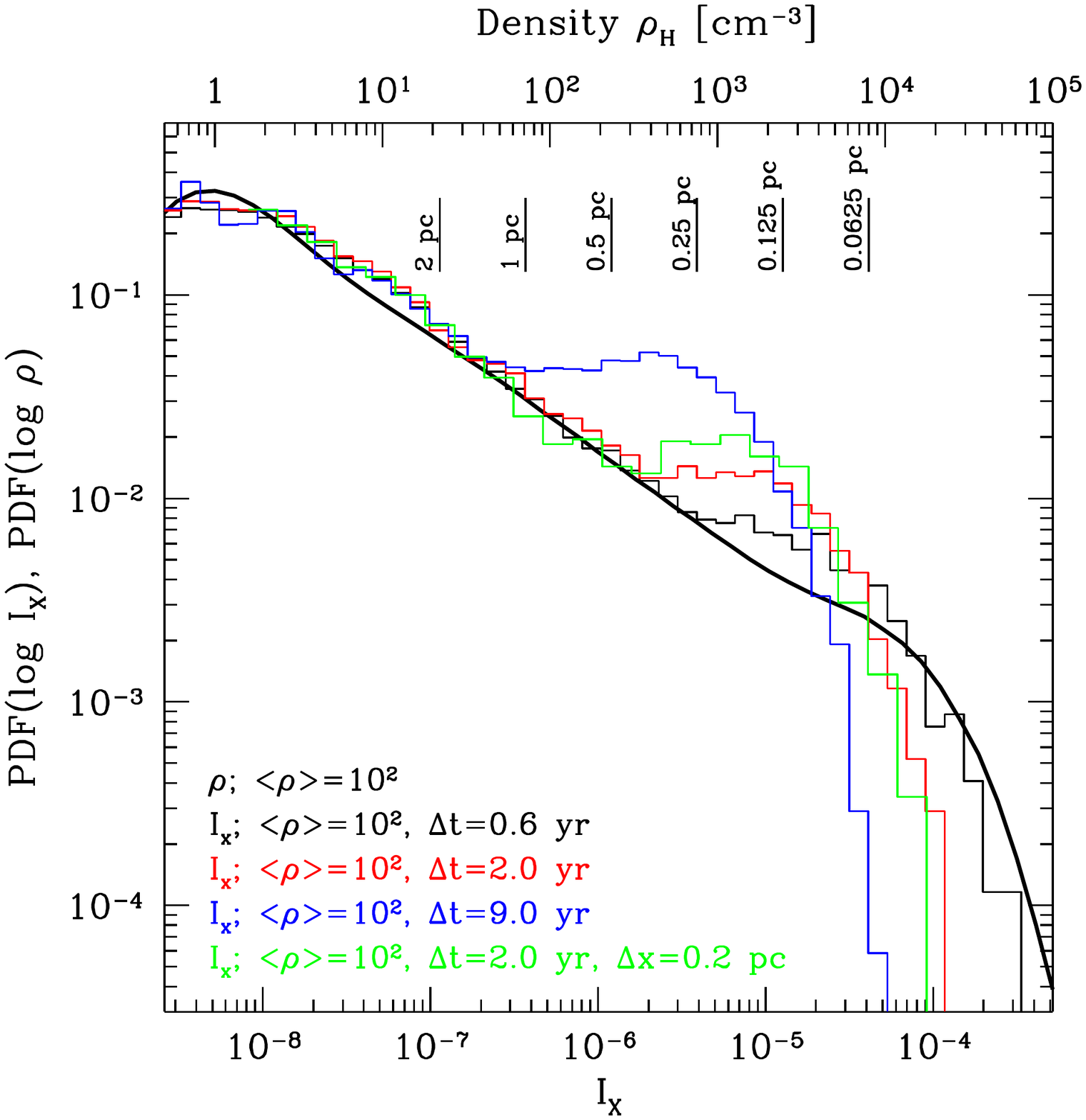}
\end{minipage}
\begin{minipage}{0.49\textwidth}
\includegraphics[trim= 0cm 4cm 0mm 2cm,width=1\textwidth,clip=t,angle=0.,scale=0.98]{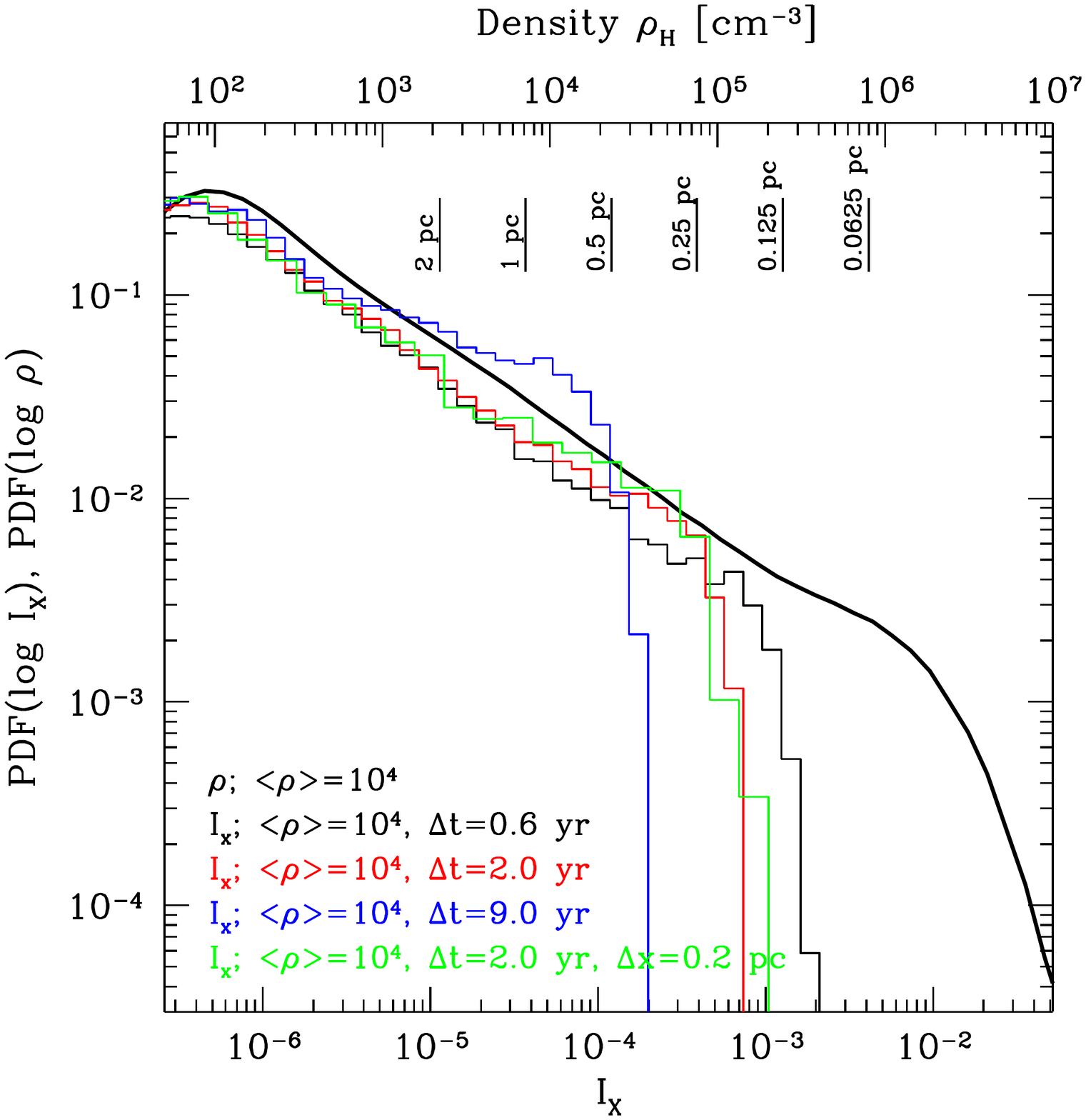}
\end{minipage}
\caption{Simulated $PDF(\rho)$ and $PDF(I_X)$ for a molecular cloud illuminated by an X-ray flare from Sgr~A*. The left and right panels correspond to the clouds with mean hydrogen densities $10^2$ and $10^4\; {\rm cm^{-3}}$, respectively. In both plots the thick black line shows the $PDF(\rho)$ ($\rho$ is shown in the upper horizontal axis), while the histograms correspond to the observed  $PDF(I_X)$ for flares with different durations ($I_X$ is shown in the lower horizontal axis). The two horizontal axes are aligned such that in the optically thin limit a given value of $\rho$ corresponds to the value of $I_X$, so that $PDF(\rho)$ and $PDF(I_X)$ can be compared directly. The cloud in the left panel is optically thin and for a short flare ($\Delta t=0.6\;{\rm yr}$ the black ``observed'' histrogram reproduces will  $PDF(\rho)$. As the duration of the flare increases, the size of the illuminated region along the line of sight $\Delta z$ becomes larger than the size of small and dense cores and the value $I_X$ moves to the left, forming a distinct bump in the  $PDF(I_X)$ (see text for details). For the denser cloud (right panel), small clumps become optically thick and a cutoff in $PDF(I_X)$ develops towards high density end. These simulations show that it is possible to recover $PDF(\rho)$ for the low density part of the distribution, provided that the duration of the flare is sufficiently small.
\label{fig:pdf_sim}
}
\end{figure*}

\subsubsection{Illustrative simulation of an inhomogeneous gas distribution}
\label{sec:sim}
To illustrate the possibility to recover the $PDF(\rho)$ we simulated the illumination of a volume filled with inhomogeneous molecular gas. To this end, we used a simplified version of the recipe from \cite{2011IAUS..270..323W} to generate a  $16\times 16\times 16$~pc cube with 0.0625~pc pixels. Namely, we first assign unit density to the whole cube. We then divide the cube into 8 cub-subes, randomly select a fraction $f=0.5$ of these sub-cubes and increase the density in the selected sub-cubes by a factor $C=3.25$. The same procedure is iteratively repeated for each sub-cube with increased density until we reach the resolution of the cube. In addition, the  density in the cube is modulated multiplicatively by an additional random field  to slightly smooth the distribution.  The resulting density field has a power law spectrum with the slope that depends on $f$ and $C$. Finally, the density is multiplied by a constant to obtain desired mean density $\langle \rho \rangle$ in the cube. 

The cube center is placed 20~pc away from the primary source in the sky plane and $\sim 7\;{\rm pc}$ behind the source along the line of sight, approximately reproducing the position of molecular complex (``Bridge'' and G0.11-0.11) studied in \citeauthor{C17a}. The illuminating source has a power law spectrum with photon index $\Gamma=2$. A small grid of models was run with the duration of the flare $\Delta t$ varying between 0.6 and 9~yr. The luminosity of the flare $L_X$ was adjusted to keep the total energy $L_X\Delta t$ constant (see \S\ref{sec:tot}). The reflected flux was calculated using a single scattering approximation, which is well suited for predicting the strength of the reflected emission (\citeauthor{C17b}).

Examples of the $PDF(\rho)$ and $PDF(I_X)$ obtained in the simulations are shown in Fig.~\ref{fig:pdf_sim} for $\langle \rho \rangle$ equal to $10^2$ and $10^4\; {\rm cm^{-3}}$ (here the density $\rho$ is the number density of hydrogen). In both plots the thick black line shows the $PDF(\rho)$ ($\rho$ is shown in the upper horizontal axis). The histograms in the same figure correspond to the predicted  $PDF(I_X)$ for flares with different durations ($I_X$ is shown in the lower horizontal axis). The upper and lower  horizontal axes are aligned such that in the optically thin limit, a given value of $\rho$ corresponds to the value of $I_X$ given by eqs.~(\ref{eq:ix_thin}) and (\ref{eq:s48}). For $\langle \rho \rangle=10^2\; {\rm cm^{-3}}$, the cloud is optically thin and for a short flare ($\Delta t=0.6\; {\rm yr}$) the black histogram $PDF(I_X)$ is an excellent proxy to the true underlying $PDF(\rho)$.

\subsubsection{Impact of the flare duration and spatial binning}
As the duration of the flare increases, the size of the illuminated region along the line of sight $\Delta z$ becomes larger than the size of dense clumps. As a result, $I_X$ drops below the predictions for the case when the density is constant over entire illuminated region and the derived values of the surface brightness for the small and dense clumps moves to the left, forming a ``bump'' in the $PDF(I_X)$. The amplitude of the bump increases as one moves to longer flares (see, e.g., blue histogram in Fig.~\ref{fig:pdf_sim}). The magnitude of this effect depends on how much mass is associated with small clumps. Indeed, in the optically thin limit the scattered flux is proportional to the total illuminated mass. Thus, the flux reflected by small  clumps is conserved, but it is smeared over by averaging and contributes to lower surface brightness bins.    In the simulations shown in Fig.~\ref{fig:pdf_sim} more mass is sitting on small scales, since $f\times C>1$, and this makes the bump very prominent. 

Qualitatively similar effect will be caused by averaging observations separated by a time longer than the light crossing time of dense clumps along the line of sight, since in individual observations  each clump will be present only in some observations. 

Similar ``migration'' of $I_X$ towards lower fluxes can be caused by limited spatial resolution, e.g., by binning the pixels in the image to achieve higher signal to noise ratio. This is illustrated by the green histogram in Fig.~\ref{fig:pdf_sim}, which shows the effect of binning the image to 0.2~pc pixels (from the original resolution of $\sim0.06\;{\rm pc}$).

\subsubsection{Impact of the finite optical depth} 
Somewhat different results are obtained for the same setup, but with higher mean density $\langle \rho \rangle=10^4\; {\rm cm^{-3}}$ (see right panel of Fig.~\ref{fig:pdf_sim}). In the simulated box, the column density increases at small scales (at each level of hierarchy the column density increases by a factor $C/2\approx1.6$). As a result,  small clumps become optically thick, their reflection efficiency per unit mass decreases, and in the distribution of $I_X$ they move to the left. Thus, a cutoff of the $PDF(I_X)$ compared to underlying $PDF(\rho)$ at large densities is the result of the opacity effects. The part of the $PDF(I_X)$, corresponding to smaller densities, is recovered well. Unlike the optically thin case, the reflected flux associated with dense clumps is not conserved and the bump is much less pronounced. In addition, if the median value of the optical depth over entire region becomes significant, the attenuation for any line of sight contributes to an overall shift of the histogram towards lower fluxes (see  right panel of Fig.~\ref{fig:pdf_sim}).

Qualitatively similar results were obtained by simulating a log-normal density distribution instead of a power law. This was done by generating many randomly positioned  3D structures (e.g. spheres) in a simulated volume. The sum of all  structures overlapping at a given position gives $\log \rho$. The mean and the width of the resulting distribution can be controlled by choosing the volume filling factor and the value assigned to each structure.  Thus generated density cube was used for simulations of the reflected emission (see \S\ref{sec:sim}).
As before, in the optically thin case the density distribution is recovered well from the distribution of $I_X$. For the case, when the opacity is significant, both effects i) disappearance of the high column density clumps and ii) overall shift of the PDF to the left due to distributed absorption are present. The magnitudes of these effects depend on the distribution of the mass across spatial scales.  When significant fraction of the mass is confined to small number of very dense clumps, only those clumps are affected by the opacity, while the rest of the $PDF(\rho)$ is preserved.

Of course, such simulations are not intended to reproduce realistic density $PDF$ of molecular clouds. They show instead, that i) it is possible to recover $PDF(\rho)$ for the low density part of the distribution, which is not affected by the opacity and ii) the short flare case is a more direct tracer of the density distribution than the case when the duration of the flare is longer than few years.  

\subsubsection{$PDF(I_X)$ in the existing \Chandra data and possible improvements}
\label{sec:pdf_obs}
\begin{figure}
\includegraphics[trim= 0mm 5cm 0cm 3cm, width=1\textwidth,clip=t,angle=0.,scale=0.48]{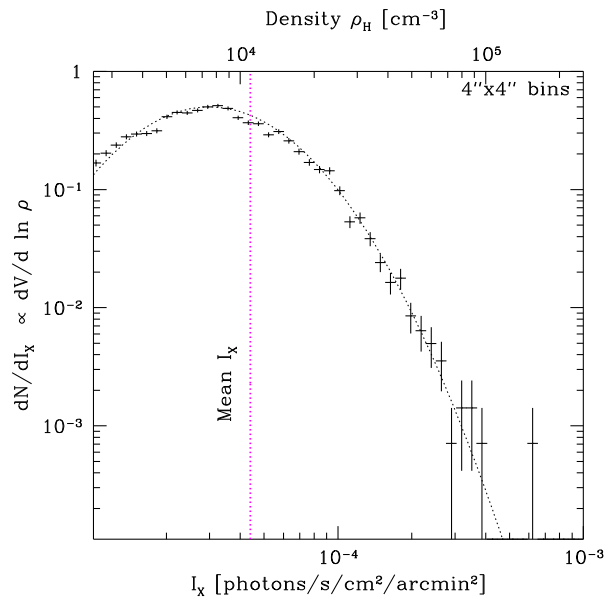}
\caption{Observed Probability Distribution Function of the X-ray surface brightness $I_X$ in the reflected component for  $4''$ bins, corresponding to $\sim0.16~{\rm pc}$. In the optically thin limit $PDF(I_X)$ is essentially $PDF(\rho)$ with proper transformation of arguments. The upper horizontal axis shows the density obtained by converting $I_X$ to the gas density, using eq.~(\ref{eq:ix_thin}) in the limit of an optically thin cloud. The decline at large densities is either due to averaging over the bin's volume, opacity effects, or the paucity of small and dense clumps. The  $PDF(I_X)$ based on the existing data is strongly affected by the photon counting noise, especially at low flux levels.
\label{fig:hib}
}
\end{figure}

Shown in Fig.~\ref{fig:hib} is the observed $PDF(I_X)$, based on \Chandra data taken in 2015 (see the right panel in Fig.~\ref{fig:tomo}). To increase statistical significance of the measured reflection component flux in individual pixels, the image was binned to $4''\times 4''$ pixes. The dotted vertical line shows the mean surface brightness over the region studied.  The dashed line shows the log-normal distribution with $\sigma_s\approx 0.7$, where $\sigma_s$ is the  standard deviation of $\log I_X$. The standard deviation $\sigma_I$ of the surface brightness normalized by the mean value, can be obtained from  $\sigma_s$ as $\displaystyle\sigma_I=\left[\exp \left(\sigma_s^2\right)-1\right]^{1/2}\sim0.8$ \citep[][]{2011ApJ...727L..21P}.  The upper horizontal axis shows the density, corresponding to a given $I_X$ in the lower horizontal axis, assuming an optical thin case and using the value $L_X\Delta t$ estimated in \S\ref{sec:tot}.  Therefore, the conversion to gas density bears the same level of uncertainties as $L_X\Delta t$.
 
While it is tempting to conclude that the density distribution is log-normal, two caveats have to be considered. First, the low flux part of the distribution (below $\sim 10^{-4}~{\rm phot s^{-1} cm^{-2}~arcmin^{-2}}$)  is  strongly dominated by the statistical noise. Secondly, while the statistical uncertainty in getting $I_X$ at high fluxes is not an issue, the measured distribution can suffer from the effects considered above, namely: opacity and time/space averaging, that could lead to a suppression of the high flux end of the distribution. 

Some of these shortcomings can be eliminated in near future. For example, ongoing \Chandra observations (PI: Maica Clavel) of the molecular clouds in the Galactic Center region should be able to constrain $\Delta t$ sufficiently well to remove the uncertainty in the duration of the flare. These observations consist of several $\sim$100~ks exposures spread over few years. Further progress is possible with an additional very deep ($\sim$ 1~Ms long) and continuous (to avoid smearing of image due to light propagation) \Chandra observation.  Such observation could help in several ways: i) by improving the significance of flux measurements in the low surface brightness regions and therefore extending the dynamic range of the $PDF(I_X)$ on the left side, ii) by allowing for finer pixels and potentially extending the dynamic range on the high flux side and iii) by making detailed spectral analysis of many small regions to better separate different components and estimating the optical depth for multiple bright regions.  This, in particular, means that the accuracy of measuring the
reflection emission intensity using the two components decomposition
method (see \S\ref{sec:flare}) can be thoroughly verified on the
smallest scales. Although the narrow band map (close to 6.4 keV) does
show a very similar morphology and a qualitatively similar $PDF$, the
statistics in the narrow band image is lower and the direct spectral
analysis based on the longer exposure would be very helpful. Fig.~\ref{fig:clouds} shows parameters of individual (isolated) clouds needed to provide at least 100 counts in the 4-8~keV band during  100~ks and 1~Ms observations. Going from 100~ks to  1~Ms exposure means, that the densest clouds can be detected if their radii exceed $\sim1''$ approximately matching the \Chandra angular resolution. For the low density clouds, the size of the cloud corresponding to 100 counts, may exceed the size of the illuminated region $\Delta z$ along the line of sight. For such clouds, in  Fig.~\ref{fig:clouds} we make a correction to the expected flux, since only a fraction of the clouds volume is illuminated. A deep and almost continuous  observations are crucial to improve the knowledge of the bright end of the PDF, corresponding to dense compact cores, since there the variability time scales are expected to be very short. On the other hand, if the faint end of the PDF is due to the lower density and extended structure, the accumulation of the exposure over several years might still be helpful.

In addition, the results of more realistic numerical simulations of molecular clouds  could be used to perform the  analysis similar to that shown in Fig.~\ref{fig:pdf_sim}. To this end we note that zoom-in numerical simulations have already reached the resolution of $\sim 0.06\;{\rm pc}$ \citep{2017arXiv170406487S}, which is fully sufficient for confronting the models with the data of the current generation of X-ray observatories.

While the caveats discussed above can affect the derived $PDF(\rho)$, it is interesting to speculate on the cloud characteristics, taking measured width of the log-normal distribution $\sigma_I\sim 0.8$ at the face value. Recently, \cite{2016ApJ...832..143F} analyzed the properties of the ``Brick'' cloud, deriving $\sigma_{\rho/\rho_0}\sim 1.3$ and concluded that the width of the density PDF suggests a predominantly solenoidal driving of the turbulence in the cloud. Since the solenoidal driving is less efficient in promoting formation of dense cores, this result might explain a low star formation efficiency in the ``Brick'' cloud as found by \cite{2016arXiv161003499K,2016arXiv161003502K}. Assuming that in the cloud currently illuminated by the Sgr~A* flare all conditions, except for the width of the PDF, are similar to those in the ``Brick'', our value of $\sigma_I$ will lead to even smaller values of turbulence driving parameter $b$ \citep[see][for the definition of this parameter]{2016ApJ...832..143F}. 
While it is clear that existing data are not yet sufficient to make a firm conclusion, the possibility of direct measurements of the density PDF for illuminated clouds looks very attractive.

\subsection{Structure function of gas density}
\label{sec:sf}
\begin{figure}
\includegraphics[trim= 0mm 6cm 0cm 3cm, width=1\textwidth,clip=t,angle=0.,scale=0.48]{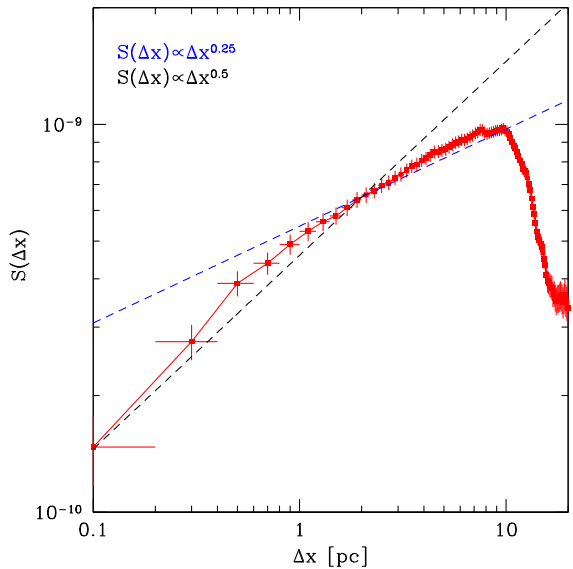}
\caption{Second order structure function $S(\Delta x)$ of the reflected emission obtained from  \Chandra observations in 2015.
The same $4''$ binned image as in Fig.~\ref{fig:hib} is used. For comparison two dashed lines show the power laws ($S\propto  \Delta x^\eta$), with $\eta=0.25$ and 0.5. In simulations \citep[see][]{2009ApJ...692..364F}, such power laws (approximately) correspond to solenoidal and compressive drivings, respectively. The observed structure function of the reflected emission intensity may be affected by the duration of the flare (mostly on small scales), by the opacity effects and by the sampling variance and the choice of analyzed region at larger scales. In particular, the drop of the structure function at largest $\Delta x$ is clearly caused by the fact that the entire cloud is encompassed by the analyzed region and the  surface brightness is small on the opposite sides of the region.
\label{fig:sf15}
}
\end{figure}

\begin{figure}
\includegraphics[trim= 0mm 6cm 0cm 3cm, width=1\textwidth,clip=t,angle=0.,scale=0.48]{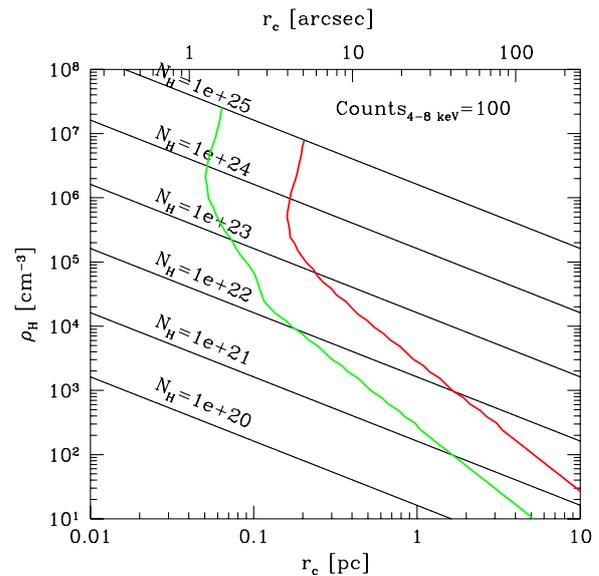}
\caption{Prospects of detecting an isolated cloud in reflected emission with \Chandra for different cloud
  sizes and densities during 100~ks (red curve) and 1~Ms (green curve)
  exposures, respectively. The size of the illuminated region along the line of sight is set to $\Delta z\approx 0.21\;{\rm pc}$
  The sensitivity is defined by the
  requirement to get 100 counts in the 4-8 keV band to spectrally
  confirm the reflection nature of the emission. 1~Ms observation could help in several ways: i) by improving the significance of flux measurements in the low surface brightness regions and therefore extending the dynamic range of the $PDF(I_X)$ on the left side (see Fig.~\ref{fig:hib}), ii) by allowing for finer pixels and potentially extending the dynamic range on the high flux side and iii) by making detailed spectral analysis of many small regions to better separate different components and estimating the optical depth for multiple bright regions.   Going from 100~ks to  1~Ms exposure means that the densest clouds can be detected if their radii exceed $\sim1''$ approximately matching \Chandra angular resolution. Further increase of the exposure will not drastically improve the prospects of finding even smaller clouds, but might still be useful to map low surface brightness regions. 
\label{fig:clouds}
}
\end{figure}

Another important characteristic of the molecular gas density distribution is the structure function. The second order structure function of the density field is defined as 
\begin{eqnarray}
S(\Delta \vec{x})=\left \langle \left [ \rho(\vec{x})-\rho(\vec{x}+\Delta \vec{x})\right ]^2\right \rangle,
\label{eq:sf}
\end{eqnarray}
where $\rho(\vec{x})$ is the density at a position $\vec{x}$. We assume that the structure function is isotropic, i.e., it depends on the absolute value of $\Delta \vec{x}$. In the optically thin limit (and a short flare), the structure function of the surface brightness can be used in lieu of the density field (see \citeauthor{C17a}). The  structure function of the reflected emission obtained from  \Chandra images taken in 2015 is shown in Fig.~\ref{fig:sf15}. The same $4''$ binned image as in Fig.~\ref{fig:hib} was used to calculate the structure function. At small $\Delta x$ it is shallower than the structure function derived from the \XMM data in \citeauthor{C17a}, due to the better angular resolution of \Chandra. The slope of $S(\Delta x)$ can be used to place constraints on the internal structure of the molecular gas. For instance,  two dashed lines show the power laws ($S\propto  \Delta x^\eta$), with $\eta=0.25$ and 0.5. In the simulation of \cite{2009ApJ...692..364F} such power laws (approximately) correspond to solenoidal and compressive drivings, respectively. As in Section \ref{sec:pdf_obs}, taking at the face value this comparison hints on larger role of the solenoidal driving. We reiterate here that the observed PDF and the structure function can suffer from the sampling variance, especially for large $\Delta x$. For instance, the drop of the structure function at largest $\Delta x$ is clearly caused by the fact that the entire cloud is encompassed by the analyzed region and the  surface brightness is small on the opposite sides of the region. At small $\Delta x$, the structure function can be sensitive to the duration of the flare, angular resolution of the telescope and the opacity of dense clumps. 

\subsection{Gas velocities from high-resolution spectroscopy}
\label{sec:future}

In this section we briefly outline new diagnostics that is difficult to do now, but will become possible with future X-ray observatories. 

High energy resolution coupled with high angular resolution will
 become available with missions like \ATHENA ~\citep{2013arXiv1306.2307N} and \LYNX  \citep[formerly X-Ray Surveyor][]{2012SPIE.8443E..16V}. For the reflected
emission this implies that using fluorescent lines one can measure 
line-of-sight velocities of the gas illuminated by Sgr~A* with the accuracy of few km/s. Presumably,
by that time, $\Delta t$, $t_{age}$ and $L_X$ [or more generally,
entire $L_X(t)$] will already be reliably known (see \S\ref{sec:dt} above). Thus, for any clump bright in the reflected emission we will know both the line-of-sight position $z$ and the line-of-sight velocity $v_z$. Such data would dramatically broaden the possibility to use Sgr~A* illumination as a diagnostic tool, as we briefly outline below.

\begin{itemize}
\item {\bf Kinematics of molecular clouds.} Since the $z$ coordinate  is effectively measured with respect to the illuminating source -- Sgr~A*, i.e., the dynamic center of the Galaxy, a pair of values  ($z$,$v_z$)  provides an important information on the molecular gas kinematics. For instance, one can immediately verify if the clouds are on the low eccentricity orbits in the gravitational potential of the central part of the Milky Way. This test can be further enhanced by using variations  of $v_z$ with $l$ and $b$. 

Furthermore, suppose that some masers can be associated with selected dense clumps, illuminated by Sgr~A*. Suppose further, that peculiar motions of these masers in the sky plane can be measured. Then for these masers/clumps we will know both accurate 3D positions and all velocity components.

\item {\bf Structure function of the gas velocity field.} On small scales one can hope to measure the structure function of the gas velocity field. From a single observation the information on the structure function in the direction perpendicular to $v_z$ can be obtained. Multiple observations, spread over few years would provide data on the $v_z$ variations along the $z$ direction. The degree of accuracy that can be achieved crucially depends on the spectral resolution of future observatories. We note here that detecting velocity broadening in unresolved clumps (rather than variations of the line centroid) may be very challenging for X-ray observatories. 

\item {\bf Combining X-ray and molecular line data.} Having $v_z(l,b,z)$ and $\rho(l,b,z)$, one can combine these data with Position-Position-Velocity (PPV) data on molecular lines to recover 3D distribution of molecular tracers. The accuracy of this procedure will depend on the amplitude of the velocity gradients and the spectral resolution of future missions. A byproduct of the above procedure is a detailed information of the molecular line fluxes from the gas with a given density, at least for clouds not affected by the absorption of either X-rays (say, 4-8 keV) or molecular lines. Of course, these data can also be used to make constraints on the excitation mechanisms of specific lines.
\end{itemize}

This is, of course, not an exhaustive list of possibilities, but just a few examples, illustrating the power of a combination of Sgr~A* flares and molecular clouds.

\section{Conclusions}
Illumination of molecular clouds in the CMZ by a powerful X-ray flare of  Sgr~A* can be used to study the properties of such flare, even although it occurred more than a hundred years ago. The same flare can be used to probe the inner structure of molecular clouds. The better we know the properties of the flare, the better are the constraints on the properties of molecular gas, and vice versa. In \S\ref{sec:road} we outline a long-term road-map of improving our knowledge of both sides of the problem (see Table~\ref{tab:sum}).

 In  \S\ref{sec:flare} we summarize the approaches that can be used to constrain the properties of the flare. Some basic characteristics can already be measured from the existing data, albeit with low accuracy. For instance, the age of the outburst and the lower limit on the luminosity of Sgr~A* during the flare can be estimated from existing \XMM and \Chandra data. The estimate of the total energy emitted by  Sgr~A* during the flare depends on the assumption that  the mean density of illuminated gas can be estimated from molecular tracers. Assuming that this estimate is correct, the duration of the flares turns out to be short, of the order of years. Further improvements in the accuracy of flare characterization is possible with additional X-ray observations, largely based on pure geometrical arguments and making more accurate evaluation of structure functions of reflected emission in time and space domains. Future X-ray polarimetric observations would also help to eliminate remaining uncertainties, in particular, testing the main assumption that the source of illuminating photons is indeed Sgr~A*.  The best chance to probe the full light-curve of the flare could come from observations of a compact and relatively isolated clouds (yet to be identified in the data). If such cloud is found and monitored over several years, then its light-curve will be a direct proxy of the flare light-curve. Knowing the light-curve, one can use this information in order to more accurately reconstruct 3D distribution of the molecular gas.
 
 If the flare was indeed very short, the illumination of molecular clouds in the CMZ opens a unique opportunity to do in-depth diagnostics of molecular gas. In particular attractive is the possibility to avoid projection effects, since the layer of illuminated gas can be very thin (smaller than a parsec).  We argue that deep X-ray observations with \Chandra can provide direct measurements of the probability distribution function of the molecular gas density $PDF(\rho)$ and the density structure function $S(\Delta x)$ on sub-pc scales, while future observatories like \ATHENA or \LYNX will be able to reconstruct the 3D velocity field of the molecular gas and to measure the velocity structure function. 

The relation between the reflected flux and the gas density is straightforward, since the X-ray flux probes directly the gas density, being weakly sensitive to its ionization state, temperature or chemical composition. This approach has only three principal limitations: i) duration of the flare, which sets the minimal scales that can be probed, ii) opacity effects, which for photons above $\sim 4\; {\rm keV}$ become severe for the hydrogen column densities above $\sim 3\;10^{23}\;{\rm cm^{-3}}$, and iii) sampling variance. It is likely that the impact of the first two effects will become clear in the near future from the ongoing \Chandra observations. The latter limitation is especially important on largest scales (few pc - 10 pc) and can only be eliminated by finding more illuminated clouds, which will allow more accurate determination of the Sgr~A*'s flare parameters, as well as more tight constraints on the properties of molecular clouds. 

While today the estimates of the flare age (see \citeauthor{C17a}) and other parameters are still very uncertain, potentially the analysis of reflected emission could become the most accurate source of information on the clouds positions relative to Sgr~A*, their kinematics and small-scale substructure. 

\section{Acknowledgements}
The results reported in this article are based in part on data
obtained from the Chandra X-ray Observatory (NASA) Data Archive, OBSIDs: 17236 and 17239. 
We acknowledge partial support by grant No. 14-22-00271 from the Russian Scientific Foundation.

\label{lastpage}


\begin{thebibliography}{99}
\bibitem[\protect\citeauthoryear{Baganoff et al.}{2003}]{2003ApJ...591..891B} Baganoff F.~K., et al., 2003, ApJ, 591, 891 

\bibitem[\protect\citeauthoryear{Barnes et al.}{2017}]{2017arXiv170403572B} Barnes A.~T., Longmore S.~N., Battersby C., Bally J., Kruijssen J.~M.~D., Henshaw J.~D., Walker D.~L., 2017, arXiv, arXiv:1704.03572 

\bibitem[\protect\citeauthoryear{Boldyrev}{2002}]{2002ApJ...569..841B} Boldyrev S., 2002, ApJ, 569, 841 

\bibitem[\protect\citeauthoryear{Burkhart, Stalpes, \& Collins}{2017}]{2017ApJ...834L...1B} Burkhart B., Stalpes K., Collins D.~C., 2017, ApJ, 834, L1 

\bibitem[\protect\citeauthoryear{Brunt, Federrath, \& Price}{2010}]{2010MNRAS.405L..56B} Brunt C.~M., Federrath C., Price D.~J., 2010, MNRAS, 405, L56 

\bibitem[\protect\citeauthoryear{Capelli et al.}{2012}]{2012A&A...545A..35C} Capelli R., Warwick R.~S., Porquet D., Gillessen S., Predehl P., 2012, A\&A, 545, A35 
\bibitem[\protect\citeauthoryear{Cho \& Kim}{2011}]{2011MNRAS.410L...8C} Cho W., Kim J., 2011, MNRAS, 410, L8 

\bibitem[\protect\citeauthoryear{Churazov et al.}{1993}]{1993A&AS...97..173C} Churazov E., et al., 1993, A\&AS, 97, 173 

\bibitem[\protect\citeauthoryear{Churazov, Sunyaev, \& Sazonov}{2002}]{2002MNRAS.330..817C} Churazov E., Sunyaev R., Sazonov S., 2002, MNRAS, 330, 817 

\bibitem[\protect\citeauthoryear{C17a}{2017a}]{C17a} Churazov E., Khabibullin I., Sunyaev R., Ponti G., 2017, MNRAS, 465, 45 
\bibitem[\protect\citeauthoryear{C17b}{2017b}]{C17b} Churazov E., Khabibullin I., Ponti G., Sunyaev R., 2017, MNRAS, 468, 165 
\bibitem[\protect\citeauthoryear{Clark et al.}{2012}]{2012MNRAS.424.2599C} Clark P.~C., Glover S.~C.~O., Klessen R.~S., Bonnell I.~A., 2012, MNRAS, 424, 2599 

\bibitem[\protect\citeauthoryear{Clavel et al.}{2013}]{2013A&A...558A..32C} Clavel M., Terrier R., Goldwurm A., Morris M.~R., Ponti G., Soldi S., Trap G., 2013, A\&A, 558, A32 

\bibitem[\protect\citeauthoryear{Clavel et al.}{2014}]{2014sf2a.conf...85C} Clavel M., Soldi S., Terrier R., Goldwurm A., Morris M.~R., Ponti G., 2014, sf2a.conf, 85
\bibitem[\protect\citeauthoryear{Couderc}{1939}]{1939AnAp....2..271C} Couderc P.
, 1939, AnAp, 2, 271

\bibitem[\protect\citeauthoryear{Elmegreen \& Scalo}{2004}]{2004ARA&A..42..211E} Elmegreen B.~G., Scalo J., 2004, ARA\&A, 42, 211 
\bibitem[\protect\citeauthoryear{Elmegreen}{2008}]{2008ApJ...672.1006E} Elmegreen B.~G., 2008, ApJ, 672, 1006-1012 

\bibitem[\protect\citeauthoryear{Federrath, Klessen, Schmidt}{2009}]{2009ApJ...692..364F} Federrath C., Klessen R.~S., Schmidt W., 2009, ApJ, 692, 364 
\bibitem[\protect\citeauthoryear{Federrath et al.}{2010}]{2010A&A...512A..81F} Federrath C., Roman-Duval J., Klessen R.~S., Schmidt W., Mac Low M.-M., 2010, A\&A, 512, A81 
\bibitem[\protect\citeauthoryear{Federrath, Klessen, \& Schmidt}{2008}]{2008ApJ...688L..79F} Federrath C., Klessen R.~S., Schmidt W., 2008, ApJ, 688, L79 
\bibitem[\protect\citeauthoryear{Federrath \& Klessen}{2013}]{2013ApJ...763...51F} Federrath C., Klessen R.~S., 2013, ApJ, 763, 51 
\bibitem[\protect\citeauthoryear{Federrath et al.}{2016}]{2016ApJ...832..143F} Federrath C., et al., 2016, ApJ, 832, 143 
\bibitem[\protect\citeauthoryear{Genzel}{1991}]{1991ASIC..342..155G} Genzel R., 1991, ASIC, 342, 155 
\bibitem[\protect\citeauthoryear{Ginsburg et al.}{2016}]{2016A&A...586A..50G} Ginsburg A., et al., 2016, A\&A, 586, A50 
\bibitem[\protect\citeauthoryear{Girichidis et al.}{2014}]{2014ApJ...781...91G} Girichidis P., Konstandin L., Whitworth A.~P., Klessen R.~S., 2014, ApJ, 781, 91 

\bibitem[\protect\citeauthoryear{Guillochon et al.}{2014}]{2014ApJ...786L..12G} Guillochon J., Loeb A., MacLeod M., Ramirez-Ruiz E., 2014, ApJ, 786, L12 
\bibitem[\protect\citeauthoryear{Hennebelle \& Chabrier}{2011}]{2011ApJ...743L..29H} Hennebelle P., Chabrier G., 2011, ApJ, 743, L29 

\bibitem[\protect\citeauthoryear{Kauffmann et al.}{2016a}]{2016arXiv161003499K} Kauffmann J., Pillai T., Zhang Q., Menten K.~M., Goldsmith P.~F., Lu X., Guzm{\'a}n A.~E., 2016, arXiv, arXiv:1610.03499 

\bibitem[\protect\citeauthoryear{Kauffmann et al.}{2016b}]{2016arXiv161003502K} Kauffmann J., Pillai T., Zhang Q., Menten K.~M., Goldsmith P.~F., Lu X., Guzm{\'a}n A.~E., Schmiedeke A., 2016, arXiv, arXiv:1610.03502 
\bibitem[\protect\citeauthoryear{Kainulainen, Federrath, \& Henning}{2014}]{2014Sci...344..183K} Kainulainen J., Federrath C., Henning T., 2014, Sci, 344, 183 
\bibitem[\protect\citeauthoryear{Klessen \& Glover}{2016}]{2016SAAS...43...85K} Klessen R.~S., Glover S.~C.~O., 2016, SAAS, 43, 85 
\bibitem[\protect\citeauthoryear{Klessen}{2000}]{2000ApJ...535..869K} Klessen R.~S., 2000, ApJ, 535, 869 
\bibitem[\protect\citeauthoryear{Kowal, Lazarian, \& Beresnyak}{2007}]{2007ApJ...658..423K} Kowal G., Lazarian A., Beresnyak A., 2007, ApJ, 658, 423 
\bibitem[\protect\citeauthoryear{Koyama et al.}{1996}]{1996PASJ...48..249K} Koyama K., Maeda Y., Sonobe T., Takeshima T., Tanaka Y., Yamauchi S., 1996, PASJ, 48, 249
\bibitem[\protect\citeauthoryear{Krivonos et al.}{2016}]{2016arXiv161203320K} Krivonos R., et al., 2016, arXiv, arXiv:1612.03320 
\bibitem[\protect\citeauthoryear{Kritsuk et al.}{2007}]{2007ApJ...665..416K} Kritsuk A.~G., Norman M.~L., Padoan P., Wagner R., 2007, ApJ, 665, 416 
\bibitem[\protect\citeauthoryear{Kritsuk, Norman, \& Wagner}{2011}]{2011ApJ...727L..20K} Kritsuk A.~G., Norman M.~L., Wagner R., 2011, ApJ, 727, L20 
\bibitem[\protect\citeauthoryear{Kruijssen et al.}{2014}]{2014MNRAS.440.3370K} Kruijssen J.~M.~D., Longmore S.~N., Elmegreen B.~G., Murray N., Bally J., Testi L., Kennicutt R.~C., 2014, MNRAS, 440, 3370 

\bibitem[\protect\citeauthoryear{Krumholz \& McKee}{2005}]{2005ApJ...630..250K} Krumholz M.~R., McKee C.~F., 2005, ApJ, 630, 250 
\bibitem[\protect\citeauthoryear{Longmore et al.}{2013}]{2013MNRAS.429..987L} Longmore S.~N., et al., 2013, MNRAS, 429, 987 

\bibitem[\protect\citeauthoryear{Marin et al.}{2015}]{2015A&A...576A..19M} Marin F., Muleri F., Soffitta P., Karas V., Kunneriath D., 2015, A\&A, 576, A19 

\bibitem[\protect\citeauthoryear{Markevitch, Sunyaev, \& Pavlinsky}{1993}]{1993Natur.364...40M} Markevitch M., Sunyaev R.~A., Pavlinsky M., 1993, Nature, 364, 40 
\bibitem[\protect\citeauthoryear{McKee \& Ostriker}{2007}]{2007ARA&A..45..565M} McKee C.~F., Ostriker E.~C., 2007, ARA\&A, 45, 565 
\bibitem[\protect\citeauthoryear{Mac Low \& Klessen}{2004}]{2004RvMP...76..125M} Mac Low M.-M., Klessen R.~S., 2004, RvMP, 76, 125 

\bibitem[\protect\citeauthoryear{Mac Low}{1999}]{1999ApJ...524..169M} Mac Low M.-M., 1999, ApJ, 524, 169 

\bibitem[\protect\citeauthoryear{Mills \& Battersby}{2017}]{2017ApJ...835...76M} Mills E.~A.~C., Battersby C., 2017, ApJ, 835, 76 
\bibitem[\protect\citeauthoryear{Molina et al.}{2012}]{2012MNRAS.423.2680M} Molina F.~Z., Glover S.~C.~O., Federrath C., Klessen R.~S., 2012, MNRAS, 423, 2680 

\bibitem[\protect\citeauthoryear{Molaro, Khatri, \& Sunyaev}{2016}]{2016A&A...589A..88M} Molaro M., Khatri R., Sunyaev R.~A., 2016, A\&A, 589, A88 

\bibitem[\protect\citeauthoryear{Mori et al.}{2015}]{2015ApJ...814...94M} Mori K., et al., 2015, ApJ, 814, 94 

\bibitem[\protect\citeauthoryear{Muno et al.}{2007}]{2007ApJ...656L..69M} Muno M.~P., Baganoff F.~K., Brandt W.~N., Park S., Morris M.~R., 2007, ApJ, 656, L69 

\bibitem[\protect\citeauthoryear{Nandra et al.}{2013}]{2013arXiv1306.2307N} Nandra K., et al., 2013, arXiv, arXiv:1306.2307 

\bibitem[\protect\citeauthoryear{Nolan, Federrath, \& Sutherland}{2015}]{2015MNRAS.451.1380N} Nolan C.~A., Federrath C., Sutherland R.~S., 2015, MNRAS, 451, 1380 

\bibitem[\protect\citeauthoryear{Ostriker, Stone, \& Gammie}{2001}]{2001ApJ...546..980O} Ostriker E.~C., Stone J.~M., Gammie C.~F., 2001, ApJ, 546, 980 
\bibitem[\protect\citeauthoryear{Ostriker, Gammie, \& Stone}{1999}]{1999ApJ...513..259O} Ostriker E.~C., Gammie C.~F., Stone J.~M., 1999, ApJ, 513, 259 

\bibitem[\protect\citeauthoryear{Padoan \& Nordlund}{2002}]{2002ApJ...576..870P} Padoan P., Nordlund {\AA}., 2002, ApJ, 576, 870 

\bibitem[\protect\citeauthoryear{Padoan, Nordlund, \& Jones}{1997}]{1997MNRAS.288..145P} Padoan P., Nordlund A., Jones B.~J.~T., 1997, MNRAS, 288, 145 
\bibitem[\protect\citeauthoryear{Padoan \& Nordlund}{2011}]{2011ApJ...730...40P} Padoan P., Nordlund {\AA}., 2011, ApJ, 730, 40 

\bibitem[\protect\citeauthoryear{Passot \& V{\'a}zquez-Semadeni}{1998}]{1998PhRvE..58.4501P} Passot T., V{\'a}zquez-Semadeni E., 1998, PhRvE, 58, 4501 

\bibitem[\protect\citeauthoryear{Ponti et al.}{2010}]{2010ApJ...714..732P} Ponti G., Terrier R., Goldwurm A., Belanger G., Trap G., 2010, ApJ, 714, 732 

\bibitem[\protect\citeauthoryear{Ponti et al.}{2013}]{2013ASSP...34..331P} Ponti G., Morris M.~R., Terrier R., Goldwurm A., 2013, ASSP, 34, 331 

\bibitem[\protect\citeauthoryear{Price, Federrath, \& Brunt}{2011}]{2011ApJ...727L..21P} Price D.~J., Federrath C., Brunt C.~M., 2011, ApJ, 727, L21 
\bibitem[\protect\citeauthoryear{Ryu et al.}{2013}]{2013PASJ...65...33R} Ryu S.~G., Nobukawa M., Nakashima S., Tsuru T.~G., Koyama K., Uchiyama H., 2013, PASJ, 65, 33 

\bibitem[\protect\citeauthoryear{Salim, Federrath, \& Kewley}{2015}]{2015ApJ...806L..36S} Salim D.~M., Federrath C., Kewley L.~J., 2015, ApJ, 806, L36 

\bibitem[\protect\citeauthoryear{Seifried et al.}{2017}]{2017arXiv170406487S} Seifried D., et al., 2017, arXiv, arXiv:1704.06487 

\bibitem[\protect\citeauthoryear{Soffitta et al.}{2013}]{2013ExA....36..523S} Soffitta P., et al., 2013, ExA, 36, 523 


\bibitem[\protect\citeauthoryear{Stone, Ostriker, \& Gammie}{1998}]{1998ApJ...508L..99S} Stone J.~M., Ostriker E.~C., Gammie C.~F., 1998, ApJ, 508, L99 
\bibitem[\protect\citeauthoryear{Stutzki et al.}{1998}]{1998A&A...336..697S} Stutzki J., Bensch F., Heithausen A., Ossenkopf V., Zielinsky M., 1998, A\&A, 336, 697 

\bibitem[\protect\citeauthoryear{Sunyaev et al.}{1991}]{1991ApJ...383L..49S} Sunyaev R., et al., 1991, ApJ, 383, L49 

\bibitem[\protect\citeauthoryear{Sunyaev, Markevitch, \& Pavlinsky}{1993}]{1993ApJ...407..606S} Sunyaev R.~A., Markevitch M., Pavlinsky M., 1993, ApJ, 407, 606 

\bibitem[\protect\citeauthoryear{Sunyaev \& Churazov}{1996}]{1996AstL...22..648S} Sunyaev R.~A., Churazov E.~M., 1996, AstL, 22, 648 

\bibitem[\protect\citeauthoryear{Sunyaev, Uskov, \& Churazov}{1999}]{1999AstL...25..199S} Sunyaev R.~A., Uskov D.~B., Churazov E.~M., 1999, AstL, 25, 199 

\bibitem[\protect\citeauthoryear{Sunyaev \& Churazov}{1998}]{1998MNRAS.297.1279S} Sunyaev R., Churazov E., 1998, MNRAS, 297, 1279 

\bibitem[\protect\citeauthoryear{Terrier et al.}{2010}]{2010ApJ...719..143T} Terrier R., et al., 2010, ApJ, 719, 143 
\bibitem[\protect\citeauthoryear{Vainshtein, Syunyaev, \& Churazov}{1998}]{1998AstL...24..271V} Vainshtein L.~A., Syunyaev R.~A., Churazov E.~M., 1998, AstL, 24, 271 
\bibitem[\protect\citeauthoryear{Vazquez-Semadeni}{1994}]{1994ApJ...423..681V} Vazquez-Semadeni E., 1994, ApJ, 423, 681 

\bibitem[\protect\citeauthoryear{Vikhlinin et al.}{2012}]{2012SPIE.8443E..16V} Vikhlinin A., et al., 2012, SPIE, 8443, 844316 

\bibitem[\protect\citeauthoryear{Walch et al.}{2011}]{2011IAUS..270..323W} Walch S., Whitworth A., Bisbas T., W{\"u}nsch R., Hubber D., 2011, IAUS, 270, 323 
\bibitem[\protect\citeauthoryear{Walch et al.}{2012}]{2012MNRAS.427..625W} Walch S.~K., Whitworth A.~P., Bisbas T., W{\''u}nsch R., Hubber D., 2012, MNRAS, 427, 625 

\bibitem[\protect\citeauthoryear{Walls et al.}{2016}]{2016MNRAS.463.2893W} Walls M., Chernyakova M., Terrier R., Goldwurm A., 2016, MNRAS, 463, 2893
\bibitem[\protect\citeauthoryear{Weisskopf et al.}{2013}]{2013SPIE.8859E..08W} Weisskopf M.~C., et al., 2013, SPIE, 8859, 885908 
\bibitem[\protect\citeauthoryear{Zhang et al.}{2015}]{2015ApJ...815..132Z} Zhang S., et al., 2015, ApJ, 815, 132 
\bibitem[\protect\citeauthoryear{Zubovas, Nayakshin, \& Markoff}{2012}]{2012MNRAS.421.1315Z} Zubovas K., Nayakshin S., Markoff S., 2012, MNRAS, 421, 1315 

\end{thebibliography}
\end{document}